\documentclass[a4paper,11pt]{article}
\pdfoutput=1 
\usepackage{jheppub} 

\usepackage[utf8]{inputenc}
\usepackage{amsmath,amssymb,amsthm}
\usepackage{hyperref}
\usepackage{bbm}
\usepackage{booktabs}

\newcommand\beqa{\begin{eqnarray}}
\newcommand\eeqa{\end{eqnarray}}

\newcommand{\la}[1]{\label{#1}}
\newcommand{\eq}[1]{(\ref{#1})}

\usepackage[utf8]{inputenc}
\usepackage{amsmath}
\usepackage{graphicx}
\usepackage{xcolor}
\usepackage{indentfirst}

\def\bP{{\bf P}}
\def\bQ{{\bf Q}}

\newcommand{\cQ}{\mathcal{Q}}
\newcommand{\cO}{\mathcal{O}}
\newcommand{\tM}{\tilde{M}}
\newcommand{\hM}{\hat{M}}
\newcommand{\tr}{{\rm tr}}
\newcommand{\Tr}{{\rm Tr}}
\newcommand{\p}{{\partial}}

\def\({\left(}
\def\){\right)}
\def\[{\left[}
\def\]{\right]}

\newcommand{\sym}{$\mathcal{N}=4$ SYM}

\newcommand{\beq}{\begin{equation}}
\newcommand{\eeq}{\end{equation}}
\newcommand{\beqq}{\begin{equation*}}
\newcommand{\eeqq}{\end{equation*}}

\title{${\cal N}=4$ SYM Quantum Spectral Curve in BFKL regime} 
\author{Mikhail Alfimov$^{1,2,3}$,}
\author{Nikolay Gromov$^{4,5}$,}
\author{and Vladimir Kazakov$^{6}$}

\date{July 2019}
\affiliation[1]{National Research University Higher School of Economics, ul. Usacheva, d. 6, Moscow, Russian Federation 119048}
\affiliation[2]{National Research Nuclear University Moscow Engineering Physical Institute, Kashirskoe sh., d. 31, Moscow, Russia 115409}
\affiliation[3]{P.N. Lebedev Physical Institute of the Russian Academy of Sciences, Leninskiy pr., d. 53, Moscow, Russia 119991}
\affiliation[4]{Mathematics Department, King's College London, The Strand, London WC2R 2LS, UK}
\affiliation[5]{St.Petersburg INP, Gatchina, 188 300, St.Petersburg, Russia}
\affiliation[6]{Laboratoire de Physique de l'\'Ecole Normale Sup\'erieure, ENS, Universit\'e PSL, CNRS, Sorbonne Universit\'e, Universit\'e de Paris, F-75005 Paris, France}

\emailAdd{malfimov@hse.ru}
\emailAdd{nikgromov@gmail.com}
\emailAdd{Vladimir.Kazakov@ens.fr}

\abstract{

We review the applications of the Quantum Spectral Curve (QSC) method to the Regge (BFKL) limit in $\mathcal{N}=4$ supersymmetric Yang-Mills theory. 
QSC, based on quantum integrability of the AdS$_5$/CFT$_4$ duality, was initially developed as a tool for the study of the spectrum of anomalous dimensions of local operators in the $\mathcal{N}=4$ SYM in the planar, $N_c\to\infty$ limit. 
We explain how to apply the QSC for the BFKL limit, which requires non-trivial analytic continuation in spin $S$ and extends the initial construction to non-local light-ray operators.
We give a brief review of high precision non-perturbative numerical solutions and analytic perturbative data resulting from this approach.
We also describe as a simple example of the QSC construction the leading order in the BFKL limit.
We show that the QSC substantially simplifies in this limit and reduces
to the Faddeev-Korchemsky Baxter equation for Q-functions.  
Finally, we review recent results for the  Fishnet CFT, which carries a number of similarities with the Lipatov's integrable spin chain for interacting reggeized gluons.
}

\begin{document}

\maketitle

\section{Introduction}\label{sec1.1}

The Balitsky-Fadin-Kuraev-Lipatov (BFKL) approximation~\cite{Kuraev:1977fs,Balitsky:1978ic} in Quantum Chromodynamics (QCD) has marked the beginning of a new era in the study of the properties of hadron collisions at high energy. Among various important results following BFKL approach, the Pomeron spectrum was calculated in the leading~\cite{Kuraev:1977fs,Balitsky:1978ic,Lipatov:1985uk}, and then, 20 years later, in the next-to-leading~\cite{Fadin:1998py} logarithmic approximations (LO and NLO). The complexity of the conventional Feynman
perturbation theory increases dramatically at each order and it is unlikely that the NNLO calculation in QCD is reachable with the currently available techniques.

The history of integrability in gauge theories starts from the renowned Lev  Lipatov's work \cite{Lipatov:1993yb}, where the effective Lagrangian for reggeized gluons in QCD was shown to be integrable. Lipatov was also the first to notice the equivalence of this system to the $SL(2,\mathbb{C})$ integrable quantum spin chain~\cite{Lipatov:1993qn}. Following the seminal paper~\cite{Faddeev:1994zg}, the integrability approach to Lipatov's model has been substantially advanced in \cite{Derkachov:2001yn,DeVega:2001pu,Derkachov:2002wz,Derkachov:2002pb}. 

The BFKL approximation was then successfully applied to the study of the spectrum of anomalous dimensions of certain operators in $\mathcal{N}=4$ SYM theory \cite{Kotikov:2000pm}. The result of this calculation appears to be
slightly less complex than the  original Regge limit in QCD,
possibly due to the super-symmetry.
As it was noticed in~\cite{Kotikov:2000pm,Kotikov:2002ab}
the result follows the so-called maximal  transcendentality selection principle -- only the most complicated in transcedentality parts of the QCD result are present in its \sym\;counterpart, with exactly the same coefficients\footnote{See also \cite{Kotikov:2007cy}.}.  
Even in this, maximally super-symmetric theory the complexity of the traditional  Feynman perturbation technique increases factorially at each order. 
However, a powerful alternative method is available in this theory:
due to the  quantum integrability of planar $\mathcal{N}=4$ SYM theory it became possible to compute the spectrum of the planar $\mathcal{N}=4$ SYM theory at any value of the 't~Hooft coupling $\lambda=16\pi^2 g^2=g_{YM}^2 N_c$! 
This integrability was discovered at one loop in the seminal Minahan-Zarembo paper~\cite{Minahan:2002ve}~\footnote{The first glimpses on such integrability can be found in L.Lipatov's talks~\cite{Lipatov:1997vu,Lipatov:2001fs}.}, then it was demonstrated at two-loops~\cite{Beisert:2003tq},
then at strong coupling in~\cite{Bena:2003wd,Kazakov:2004qf,Beisert:2005bm}, via AdS/CFT correspondence, 
 then generalized up to wrapping order in ~\cite{Janik:2006dc,Beisert:2005if,Beisert:2006ez,Ambjorn:2005wa}, and finally to all loops, first in terms of the exact Y-system~\cite{Gromov:2009bc}, and then of the TBA equations~\cite{Bombardelli:2009ns,Gromov:2009tv,Arutyunov:2009ur} for the simplest Konishi-like operators.
A few years ago, this line of research resulted in the final formulation of the solution of the spectral problem  -- the Quantum Spectral Curve (QSC) of the AdS$_5$/CFT$_4$ duality~\cite{Gromov:2014caa,Gromov:2015dfa}.  

The QSC equations can be formulated as a finite system of Riemann-Hilbert-type equations on 4+4 Baxter Q-functions. The QSC formalism has led to the tremendous progress in high precision numerical and multi-loop computations in  a multitude of particular physical quantities~(see the reviews \cite{Gromov:2017blm,Kazakov:2018hrh} and citations therein). Notably, the dimension $\Delta$ of twist-2 operator of the type $\tr(Z\nabla_+^S Z)$ was computed for virtually arbitrary coupling and arbitrary complex values of spin  $S$~\cite{Gromov:2015wca}~(see the Figure \ref{operator_trajectories_nonzero_conformal_spin} and Figure \ref{Riemann_surface_twist_2_operators}). The particular corner of this picture, when $S \simeq -1+{\cal O}(g^2)$ at weak coupling $g^2 \to 0$, corresponds to the BFKL regime. The loop expansion of $S$ can be written as follows
\beq\label{S_exp_BFKL}
    S=-1+g^2 \chi_{\textrm{LO}}(\Delta)+g^4 \chi_{\textrm{NLO}}(\Delta)+g^6 \chi_{\textrm{NNLO}}(\Delta)+g^8 \chi_{\textrm{NNNLO}}(\Delta)+\mathcal{O}(g^8)\;,
\eeq
where $\chi_{\textrm{N}^k \textrm{LO}}(\Delta)$ is the BFKL kernel eigenvalue at the $(k+1)$-st loop. The  BFKL limit of QSC was first explored in~\cite{Alfimov:2014bwa} and the LO BFKL spectrum $\chi_{\textrm{LO}}(\Delta)$ with zero conformal spin was reproduced there. The coefficient $\chi_{\textrm{NLO}}(\Delta)$ calculated in \cite{Kotikov:2000pm} was successfully reproduced by the QSC calculations in \cite{Gromov:2015vua}. Then, significant progress was made in the same work \cite{Gromov:2015vua}, where QSC allowed to derive the NNLO BFKL kernel eigenvalue $\chi_{\textrm{NNLO}}(\Delta)$, which was later confirmed by the other method in \cite{Caron-Huot:2016tzz}. It should be noted that all the BFKL kernel eigenvalues up to NNLO included can be written in terms of nested harmonic sums and multiple zeta values. Moreover,  the high precision QSC numerical technique~\cite{Gromov:2015wca} is currently adapted in~\cite{Alfimov:2018cms}: for any order $\chi_{\textrm{N}^k \textrm{LO}}(\Delta)$\footnote{In the case of non-zero conformal spin some data connected to the NNNLO BFKL kernel eigenvalues were extracted from QSC in \cite{Alfimov:2018cms}, namely: the intercept function (BFKL kernel eigenvalue at $\Delta=0$) for different values of conformal spin, slope-to-intercept function (BFKL kernel eigenvalue derivative with respect to conformal spin) and curvature function (BFKL kernel eigenvalue 2nd derivative with respect to $\Delta$), both calculated at the special BPS point of dimension 0 and conformal spin 1.} can be analyzed by the QSC numerical algorithm. Moreover, the QSC approach allowed to compute anomalous dimensions of the length-2 operators with both non-zero spins, including the conformal one $n$, $\tr(Z\nabla_+^{S} \nabla_{\perp}^{n} Z)$~\cite{Alfimov:2018cms}. 

In this short review, we give a concise formulation of the QSC and its  reduction in the case of BFKL limit. We will briefly review the ideas of numerical and perturbative approaches to the solution of the QSC equations. We also shortly describe the most important numerical and perturbative analytic results for the BFKL limit of $\mathcal{N}=4$ SYM theory. 

The BFKL also gave an inspiration for a number of other research directions over the years. In particular, we will explain here how the BFKL limit influenced the currently being actively developed so-called {\it fishnet} conformal field theory (FCFT) proposed in~\cite{Gurdogan:2015csr} as a double scaling limit combining weak coupling and strong $gamma$-deformation of \sym\; theory.  Fishnet CFT attracted much of attention in the last few years~\cite{Caetano:2016ydc,Chicherin:2017cns,Gromov:2017cja,Chicherin:2017frs,Grabner:2017pgm,Mamroud:2017uyz,Kazakov:2018qez,Karananas:2019fox,Gromov:2018hut,Derkachov:2018rot,Kazakov:2018gcy,Basso:2018cvy,Korchemsky:2018hnb,Chowdhury:2019hns,Gromov:2019jfh,Basso:2019xay,Derkachov:2019tzo,Adamo:2019lor,deMelloKoch:2019ywq,Pittelli:2019ceq}. It was generalized to any dimension $D$~\cite{Kazakov:2018qez}, and for $D=2$ and a specific representation for conformal spins this model can be identified (for certain physical quantities) with the Lipatov's  $SL(2,\mathbb{R})$ spin chain for LO approximation of BFKL.  As Fishnet CFT has a significant interplay with the initial ideas of BFKL integrability, we briefly review here the construction and integrability properties of fishnet CFT.

\section{Integrability in \texorpdfstring{$\mathcal{N}=4$}{\mathcal{N}=4} SYM}

In the present Section we briefly review the integrable structure of \sym. We connect the BFKL kernel eigenvalues to the dimensions of the twist-2 operators following~\cite{Kotikov:2000pm}, which allows us to apply the QSC to the study of the BFKL spectrum. This step requires certain generalisation of the gluing condition in QSC for the local operators, allowing for non-integer spins. As an example we demonstrate how to reproduce the Baxter equation, which initially appeared in the problem of diagonalization of the LO BFKL kernal.

\subsection{BFKL spectrum and maximal transcendentality principle}\label{BFKL_max_transcend}

Let us start from briefly describing the relation of the BFKL spectrum in \sym\ to those in QCD. 
The key statement governing this connection is the so-called principle of maximal transcendentality \cite{Kotikov:2000pm,Kotikov:2002ab}. 
One can find a detailed description of this principle in \cite{Kotikov:2019xyg}. Below  we demonstrate this principle in application to the BFKL kernel eigenvalues. 
In what follows we use the extension of the formula \eqref{S_exp_BFKL} for the case of non-zero conformal spin $n$
\beqa
S&=&-1+g^2 \chi_{\textrm{LO}}(\Delta,n)+g^4 \chi_{\textrm{NLO}}(\Delta,n)+ \\
&+&g^6 \chi_{\textrm{NNLO}}(\Delta,n)+g^8 \chi_{\textrm{NNNLO}}(\Delta,n)+\mathcal{O}(g^{10})\;. \notag
\eeqa

Consider first the LO BFKL Pomeron kernel eigenvalues for QCD and \sym\, which can be found in \cite{Kuraev:1977fs,Balitsky:1978ic} and \cite{Kotikov:2000pm} respectively. We see that the answers are identical and are given by
\beq\la{lon}
\chi_{\textrm{LO}}(\Delta,n)=-4\left(\psi\left(\frac{1+n+\Delta}{2}\right)+\psi\left(\frac{1+n-\Delta}{2}\right)-2\psi(1)\right)\;,
\eeq
where $\Delta$ is the dimension of the state and $n$ is the conformal spin. Moreover, in the LO the BFKL kernel eigenvalues coincide for all 4D gauge theories \cite{Kotikov:2000pm}.

For the next order in perturbation theory the situation is more subtle. The NLO BFKL kernel eigenvalue\footnote{Our identification of the variable $\gamma$ from \cite{Kotikov:2000pm} with our parameters is: $\gamma=(\Delta-S)/2$, where $\Delta$ is the dimension and $S$ is the spin.} for QCD was calculated in \cite{Fadin:1998py}
\beqa\label{NLOoriginal_QCD}
& &\chi_{\textrm{NLO}}^{\textrm{QCD}}(\Delta,n)=4\left[-2\Phi\left(n,\frac{1-\Delta}{2}\right)-2\Phi\left(n,\frac{1+\Delta}{2}\right)+6\zeta(3)+\right. \\
&+&\left(\frac{67}{9}-2\zeta(2)-\frac{10n_f}{9N_c}\right)\frac{\chi_{\textrm{LO}}(\Delta,n)}{4}+\psi''\left(\frac{1+n-\Delta}{2}\right)+\psi''\left(\frac{1+n+\Delta}{2}\right)- \notag \\
&-&\frac{1}{2}\left(\frac{11}{3}-\frac{2}{3}\frac{n_f}{N_c}\right)\left(\frac{\chi_{\textrm{LO}}^2(\Delta,n)}{16}-\psi'\left(\frac{1+n+\Delta}{2}\right)+\psi'\left(\frac{1+n-\Delta}{2}\right)\right)+ \notag \\
&+&\frac{\pi^2 \cos\frac{\pi(1+\Delta)}{2}}{\Delta\sin^2\frac{\pi(1+\Delta)}{2}}\left.\left(\left(3+\left(1+\frac{n_f}{N_c^3}\right)\frac{3\Delta^2-11}{4(\Delta^2-4)}\right)\delta^0_n-\left(1+\frac{n_f}{N_c^3}\right)\frac{\Delta^2-1}{8(\Delta^2-4)}\delta^2_n\right)\right]\;, \notag
\eeqa
where $N_c$ is the number of colors and $n_f$ is the number of flavours of the fermions in the considered version of QCD. The same quantity in \sym\ was first presented in \cite{Kotikov:2000pm}. We write it down here from \cite{Kotikov:2002ab}
\beqa\label{NLOoriginal_SYM}
\chi_{\textrm{NLO}}^{\mathcal{N}=4}(\Delta,n)&=&4\left[-2\Phi\left(n,\frac{1-\Delta}{2}\right)-2\Phi\left(n,\frac{1+\Delta}{2}\right)+6\zeta(3)+\right. \\
&+&\left. \frac{1}{4}\left(\frac{1}{3}-2\zeta(2)\right)\chi_{\textrm{LO}}(\Delta,n)+\psi''\left(\frac{1+n-\Delta}{2}\right)+\psi''\left(\frac{1+n+\Delta}{2}\right)\right]\;. \notag
\eeqa
The function $\Phi(n,x)$ in the eigenvalues \eqref{NLOoriginal_QCD} and \eqref{NLOoriginal_SYM} is equal to
\beqa
\Phi(n,x)&=&\sum\limits_{k=0}^\infty\frac{(-1)^{k+1}}{k+x+n/2}\left[\psi'(k+n+1)-\psi'(k+1)+(-1)^{k+1}(\beta'(k+n+1)+\right. \notag \\
&+&\left.\beta'(k+1))+\frac{1}{k+x+n/2}\left(\psi(k+n+1)-\psi(k+1)\right)\right]
\eeqa
and 
\beq
\beta'(z)=\frac{1}{4}\left[\psi'\left(\frac{z+1}{2}\right)-\psi'\left(\frac{z}{2}\right)\right]=\sum_{k=0}^{+\infty}\frac{(-1)^{k+1}}{(z+k)^2}\;.
\eeq
Now we are going to explain the maximal transcendentality principle with the example of NLO BFKL kernel eigenvalue. There is a way to rewrite both $\chi_{\textrm{LO}}(\Delta,n)$ \eqref{lon} and $\chi_{\textrm{NLO}}(\Delta,n)$ for both QCD \eqref{NLOoriginal_SYM} and \sym\ \eqref{NLOoriginal_SYM} in terms of nested harmonic sums. In the LO we get simple result
\beq\label{LO_BFKL_eigenvalue_harmonic_sums}
\chi_{\textrm{LO}}(\Delta,n)=-4\left(S_1\left(\frac{1+\Delta+n}{2}-1\right)+S_1\left(\frac{1+n-\Delta}{2}-1\right)\right)\;,
\eeq
from which we see that the LO BFKL kernel eigenvalues in QCD and \sym\ have transcendentality $1$, where the transcendentality of a harmonic sum $S_{n_1,\dots,n_m}$ is defined to be $|n_1|+\dots+|n_m|$,
transcendentality of a product is given by a sum of the transcendentalities of the multipliers. In the same way one defines the transcendentality of the MZV $\zeta_{n_1,\dots,n_m}$, which is given by the sum $n_1+\dots+n_m$. In particular, $\pi$ and $\log 2$ have transcedentality $1$.

To express \eqref{NLOoriginal_QCD} and \eqref{NLOoriginal_SYM} through nested harmonic sums we need to introduce some additional formulas. In \cite{Costa:2012cb} it was shown, that for $n=0$ we can represent the NLO BFKL kernel eigenvalue for \sym\ as follows
\beq\label{NLO_BFKL_eigenvalue_zero_n}
\chi_{\textrm{NLO}}^{\mathcal{N}=4}(\Delta,0)=F_2\left(\frac{1+\Delta}{2}\right)+F_2\left(\frac{1-\Delta}{2}\right)
\;,
\eeq
where
\beqa\label{F2_def}
F_2(x)&=&4\left(-\frac{3}{2}\zeta(3)+\pi^2 \log 2+\frac{\pi^2}{3}S_1(x-1)+\right. \\
&+&\left. \pi^2 S_{-1}(x-1)+2S_3(x-1)-4S_{-2,1}(x-1)\right)\;. \notag
\eeqa
Utilizing \eqref{NLO_BFKL_eigenvalue_zero_n} and \eqref{F2_def}, we are able to rewrite \eqref{NLOoriginal_SYM} in the following way
\beqa\label{NLO_BFKL_SYM_harmonic_sums}
\chi_{\textrm{NLO}}^{\mathcal{N}=4}(\Delta,n)&=&\frac{1}{2}\left(F_2\left(\frac{1+\Delta+n}{2}\right)+F_2\left(\frac{1-\Delta-n}{2}\right)+F_2\left(\frac{1+\Delta-n}{2}\right)+\right. \notag \\
&+&\left. F_2\left(\frac{1-\Delta+n}{2}\right)\right)+4\left(R_n\left(\frac{1-\Delta}{2}\right)+R_n\left(\frac{1+\Delta}{2}\right)\right)\;,
\eeqa
where
\beq\label{Rn_def}
R_n(\gamma)=-8\left(S_{-2}\left(\gamma+\frac{n}{2}-1\right)+\frac{\pi^2}{12}\right)\left(S_1\left(\gamma+\frac{n}{2}-1\right)-S_1\left(\gamma-\frac{n}{2}-1\right)\right)\;.
\eeq
Therefore, from \eqref{LO_BFKL_eigenvalue_harmonic_sums}, \eqref{NLO_BFKL_eigenvalue_zero_n}, \eqref{F2_def}, \eqref{NLO_BFKL_SYM_harmonic_sums} and \eqref{Rn_def} we see that all the contributions to \eqref{NLO_BFKL_SYM_harmonic_sums} have transcendentality $3$.
Furthermore, we see that all the terms of the QCD result with the maximal transcendentality
$3$ coincide exactly with the result of \sym\ !

In the next part we will explain the relation of the BFKL kernel eigenvalues to the spectrum of length-2 (twist-2 in the case of zero conformal spin) operators in \sym. Then
we will explain how to adapt the QSC for the local operators to describe non-integer spin and conformal spin, which would then lead to the BFKL kernel eigenvalue.

\subsection{Analytic structure of the twist-2 operator anomalous dimensions}\label{dim_analit_cont}

In this part we describe the analytic structure of the length-2 operators with conformal spin in \sym. Namely, we restrict ourselves to the states with the quantum numbers $J_1=2$, $J_2=J_3=0$. 
Let us first consider the case of zero conformal spin $S_2=0$, which are called the twist-2 $\mathfrak{sl}(2)$ operators
\beq\label{twist_2_operator}
\cO=\textrm{tr}Z D_+^S Z+(\textrm{permutations})\;.
\eeq
In the gauge theory the physical operators should have even number of derivatives to be conformal primaries. 
The dimensions of these physical operators $\Delta$ were calculated for arbitrary even integer $S$ up to several loops order \cite{Lipatov:2001fs,Kotikov:2002ab,Kotikov:2003fb}.
In \cite{Gromov:2013pga} it was understood how to describe these operators with integer spins within the QSC approach to the leading order in the perturbation theory.
This technique was then generalised to the higher orders in \cite{Marboe:2014gma,Marboe:2014sya,Marboe:2016igj,Marboe:2017dmb,Marboe:2018ugv}.

\begin{figure}
\centerline{
\includegraphics[width=0.7\linewidth]{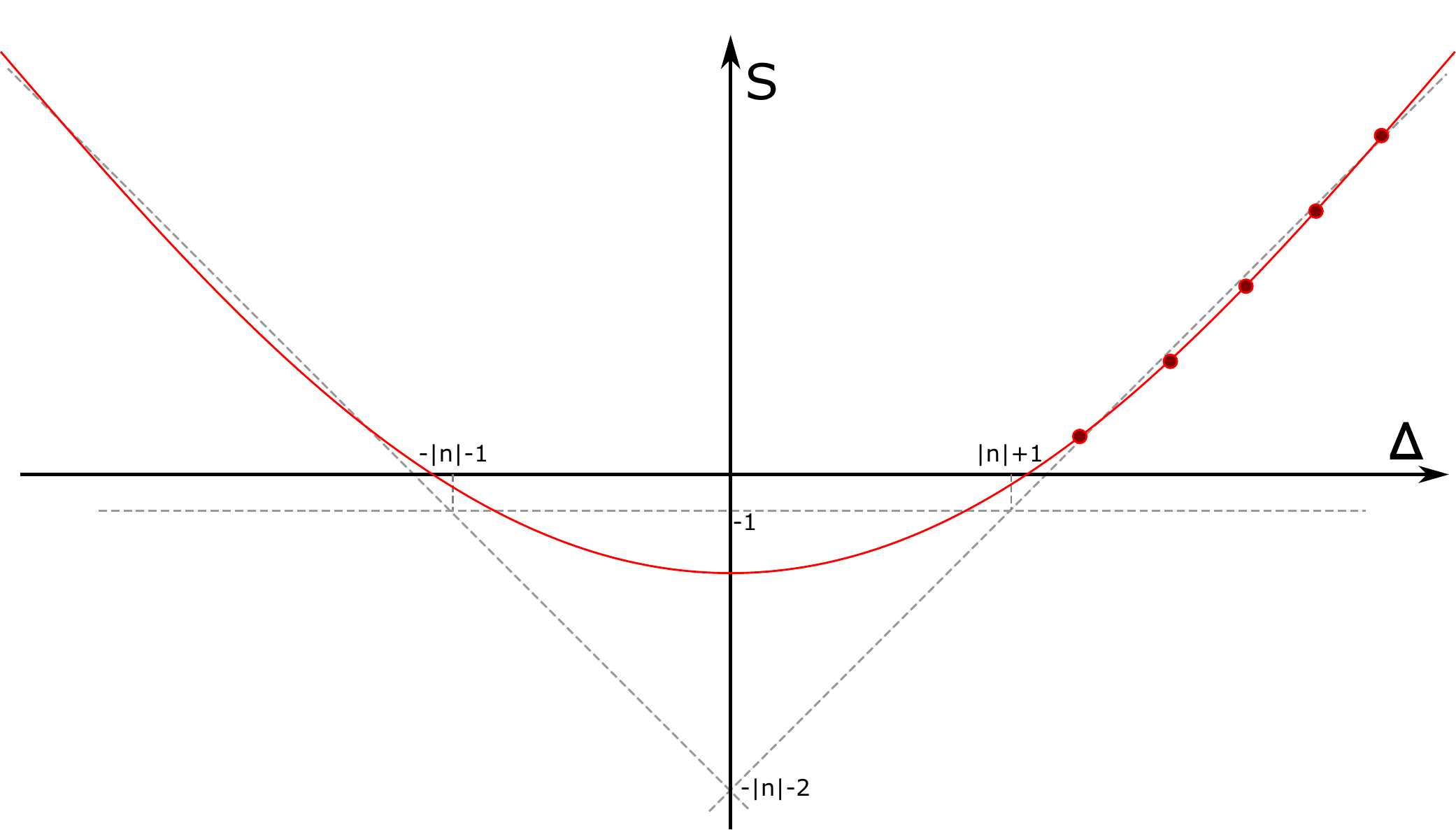}}
\caption{Trajectory  of the length-2 operator for conformal spin $n=S_2$ as a function of the full dimension $\Delta$. The dots correspond to the physical operators with $S+n \in 2\mathbb{Z}_{\geq 0}$.}
\label{operator_trajectories_nonzero_conformal_spin}
\end{figure}

\begin{figure}[ht]
\centerline{
\includegraphics[width=0.7\linewidth]{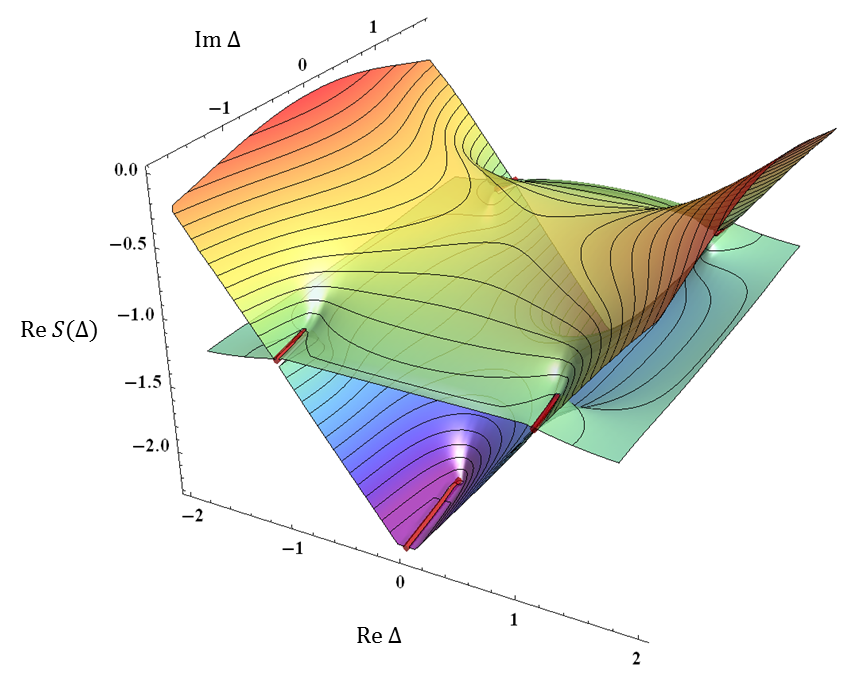}}
\caption{Riemann surface of the function $S(\Delta)$ for twist-2 operators.}
\label{Riemann_surface_twist_2_operators}
\end{figure}

As we explain below in order to make a connection with the Regge regime one needs to be able to analytically continue in $S$. In the works \cite{GromovKazakovunpublished,Janik:2013nqa} it was understood how to remove the constraint of integer even spin $S$ from the integrable description, leading to the analytic continuation of the anomalous dimension of the twist-2 $\mathfrak{sl}(2)$ operators. At the 1-loop order one gets the following result~\cite{Kotikov:2002ab,Kotikov:2005gr}
\beq\label{twist_2_1loop}
\Delta=2+S+8g^2\left(\psi(S+1)-\psi(1)\right)+\mathcal{O}(g^2)\;.
\eeq
As we can see the formula above \eqref{twist_2_1loop} is regular for all $S>-1$, but has a simple pole at $S=-1$. In fact as now both $S$ and $\Delta$ are allowed to be non-integer we can invert the function for the range where it is regular and instead consider $S(\Delta)$ which is a more convenient function \cite{Kotikov:2002ab,Brower:2006ea,Costa:2012cb,Costa:2013zra,Brower:2013jga}. In particular \eqref{twist_2_1loop} leads to
\beq\la{SDD}
S_{\Delta>1}(\Delta)=\Delta-2-
8g^2\left(\psi(\Delta-1)-\psi(1)\right)+\mathcal{O}(g^4)\;.
\eeq
Below we describe the method which would allow us to compute this function in \sym\ at finite coupling (see the Figure \ref{operator_trajectories_nonzero_conformal_spin} for $n=0$). At the moment we only need to know its main features. We see that $S(\Delta)$ is an even function of $\Delta$. At finite $g$ it has a smooth parabolic shape, however, at weak coupling 
it becomes piece-wise linear:
$S=|\Delta|-2$ for $|\Delta|>1$
and $S=-1$ for $|\Delta|<1$.
Thus the $g\to 0$ limit breaks the analyticity and analytic continuation in $S$ does not commute with the $g\to 0$ limit.
This implies that even though $S(\Delta)$ is a smooth function at finite $g$, it generates two different series expansions in $g$, depending on the range of $\Delta$. Within this picture the BFKL expansion is the series expansion of $S(\Delta)$ for $|\Delta|<1$ in powers of $g^2$.
Note that all physical operators belong to the opposite region $\Delta>1$ (the physical operator with the smallest protected dimension $2$ is $\tr(Z^2)$), and thus the equation \eq{SDD} is only valid for $\Delta>1$. At the same time for $|\Delta|<1$ one can use \eqref{lon} to get
\beq\label{LO_BFKL_eigenvalue}
S_{|\Delta|<1}(\Delta)=-1+4g^2 \left(-\psi\left(\frac{1+\Delta}{2}\right)-\psi\left(\frac{1-\Delta}{2}\right)+2\psi(1)\right)+\mathcal{O}(g^4)\;.
\eeq
To understand how \eq{SDD} and \eq{LO_BFKL_eigenvalue} are related to each other we have to complexify $\Delta$ and $S$. We explain below how this can be done using QSC at finite $g$, and the resulting function ${\rm Re}\;S(\Delta)$ is presented at the Figure \ref{Riemann_surface_twist_2_operators}.
We see that the function $S(\Delta)$ has several sheets, which are connected through a branch cut. When coupling goes to zero the branch points collide forming the singularity in the function $S(\Delta)$.
The domains $|\Delta|<1$ and $\Delta>1$ become separated completely by the branch cut dividing the $\Delta$ plane into two parts. Thus one can say that \eq{SDD} and \eq{LO_BFKL_eigenvalue} are the expansions of the two branches of the same function to the left and to the right from the quadratic cut going to infinity along imaginary axis. This observation resolves the seeming paradox that the two functions are not the same.
It also allows to make a prediction: the sum $S_{|\Delta|<1}(\Delta)+S_{\Delta>1}(\Delta)$ should be regular around $\Delta=1$ at any order in $g$, as in this combination the branch cut cancels.
Indeed, one can check that the simple poles in the both functions around $\Delta=1$ has residues opposite in sign and disappear in the sum at order $g^2$. At the higher orders the poles at $\Delta=1$ become more and more severe, but this cancellation also can be verified explicitly to all known orders.
This requirement gives non-trivial relations between the functions $S_{\Delta>1}(\Delta)$, obtained perturbatively as an analytic continuation from the dimensions of physical operators, and the BFKL kernel eigenvalue $S_{|\Delta|<1}(\Delta)$.

Now we are ready to turn to non-zero conformal spin. Namely, non-zero conformal spin adds the derivative in the orthogonal direction to the operators \eqref{twist_2_operator}
\beq\label{length_2_operator}
\cO=\textrm{tr}Z D_+^{S_1} \partial_{\perp}^{S_2} Z+(\textrm{permutations})\;.
\eeq
The physical states now correspond to non-negative integer $S_1$ and $S_2$, whose sum is even. We follow the same strategy for \eqref{length_2_operator} as for the case of zero conformal spin. Analogously, having the anomalous dimensions for the physical operators, we can build the analytic continuation in the spins $S_1$ and $S_2$ and identify them with the spin $S$ and conformal spin $n$ respectively. This analytic continuation is illustrated with the Figure \ref{operator_trajectories_nonzero_conformal_spin}. The physical operators are designated with the dots on the operator trajectory. As in the case of zero conformal spin exchange of the roles of $\Delta$ and $S=S_1$ allows us to reach the BFKL regime.

Having these  analytic properties in mind, we understand how to apply the QSC to the study of the BFKL spectrum: we need to analytically continue to non-integer spin $S$ and conformal spin $n$ not only the anomalous dimensions, but the QSC itself: Q-functions, their asymptotics and analytic structure etc. 
In the next section we are going to start from the brief description of the QSC basics. Then we will use this setup to consider the calculation of the LO BFKL kernel eigenvalue with non-zero conformal spin by the QSC method.

\subsection{QSC approach to the BFKL spectrum of \texorpdfstring{$\mathcal{N}=4$}{\mathcal{N}=4} SYM}\label{QSC_approach_BFKL}

We are going to present here the formulation of the QSC in terms of the Q-system and gluing conditions, the details of which can be found in \cite{Gromov:2014caa,Gromov:2017blm,Alfimov:2018cms}.
Algebraic part of the QSC framework is described as follows. For the \sym\ we have the system of $2^8$ Q-functions, which are denoted as
\beq\label{Q_functions}
Q_{a_1,\ldots,a_n|i_1,\ldots,i_m}(u)\;, \quad 1 \leq n,m \leq 4.
\eeq
The Q-functions \eqref{Q_functions} have two groups of indices: $a$'s are called ``bosonic'' and $i$'s are
called ``fermionic''. They are antisymmetric with respect to the exchange of any pair of indices in these two groups. Not all of the 256 Q-functions in question are independent. They are subject to the set of the so-called Pl\"ucker's QQ-relations, which are written in \cite{Gromov:2014caa}
\beqa\label{QQ_relations}
Q_{A|I}Q_{Aab|I}&=&Q_{Aa|I}^+ Q_{Ab|I}^- - Q_{Aa|I}^- Q_{Ab|I}^+\;, \\
Q_{A|I}Q_{A|Iij}&=&Q_{A|Ii}^+ Q_{A|Ij}^- - Q_{Aa|Ii}^- Q_{Ab|Ij}^+\;, \notag \\
Q_{Aa|I}Q_{A|Ii}&=&Q_{Aa|Ii}^+ Q_{Ab|I}^- - Q_{Aa|Ii}^- Q_{A|I}^+\;, \notag
\eeqa
where $A$ and $I$ are multi-indices from the set $\{1,2,3,4\}$.

The standard normalization is chosen to be
\beq
Q_{\emptyset|\emptyset}=1\;.
\eeq
Then the structure of the QQ-relations allows to express the whole Q-system in terms of the ``basic'' set of $8$ Q-functions $Q_{a|\emptyset}$, $a=1,\ldots,4$ and $Q_{\emptyset|i}$, $i=1,\ldots,4$.

Let us now describe two symmetries of the Q-system, which respect the QQ-relations \eqref{QQ_relations}.
\begin{itemize}
\item Imposing the so-called unimodularity condition
\beq\label{unimodularity_condition}
Q_{1234|1234}=1\;,
\eeq
we are able to introduce the system of Hodge-dual Q-functions
\beq
Q^{a_1 \ldots a_n|i_1 \ldots i_m} \equiv (-1)^{nm}\epsilon^{b_{n+1} \ldots b_4 a_1 \ldots a_n}\epsilon^{j_{m+1} \ldots j_4 i_1 \ldots i_m}Q_{b_{n+1} \ldots b_4|j_{m+1} \ldots j_4}\;,
\eeq
where there is no summation over the repeated indices and which satisfy the same QQ-relations \eqref{QQ_relations}. The condition \eqref{unimodularity_condition} allows to write down the following relations between the Q-functions with the lower and upper indices
\beqa
\label{Q_up_down}
Q^{a|i}Q_{b|j}=-\delta^i_j\;,& \quad &Q^{a|i}Q_{b|i}=-\delta^a_b\;, \\
Q^{a|\emptyset}=(Q^{a|i})^+ Q_{\emptyset|i}\;,& \quad &Q^{\emptyset|i}=(Q^{a|i})^+ Q_{a|\emptyset}\;, \notag \\
Q_{a|\emptyset}Q^{a|\emptyset}=0\;,& \quad &Q_{\emptyset|i}Q^{\emptyset|i}=0\;. \notag
\eeqa

\item Another symmetry is called the H-symmetry. Its general form is given by
\beq
Q_{A|I} \rightarrow \sum\limits_{|B|=|A|,|J|=|I|}(H_b^{[|A|-|I|]})_A^B (H_f^{[|A|-|I|]})_I^J Q_{B|J}\;,
\eeq
where the sum goes over the repeated multi-indices. The definition of $H_I^J$ is the following: $(H_f)_I^J \equiv (H_F(u))_{i_1}^{j_1} (H_F(u))_{i_2}^{j_2} \ldots (H_F(u))_{i_{|I|}}^{j_{|I|}}$ and the same for $(H_b)_A^B$ and $(H_B(u))_{a_k}^{b_k}$, $k=1,\ldots,|A|$ with $H_{B,F}(u)$ being $4 \times 4$ $i$-periodic matrices. The unimodularity condition leads us to the restriction
\beq
\det H_B(u) \det H_F(u)=1\;.
\eeq

\end{itemize}

To proceed with the calculation of the spectrum of \sym\ we have to endow the described Q-system with the analytic structure. To start with, we designate the basic Q-functions with this structure as $\bP_a$ (same as $Q_{a|\emptyset}$), $\bP^a$ ($Q^{a|\emptyset}$), $\bQ_i$ ($Q_{\emptyset|i}$) and $\bQ^i$ ($Q^{\emptyset|i}$). One can find the asymptotics of the Q-functions in \cite{Gromov:2017blm}
\beq\label{PQ_asymptotics}
\bP_a \simeq A_a u^{-\tilde{M}_a}\;, \quad \bP^a \simeq A^a u^{\tilde{M}_a-1}\;, \quad \bQ_i \simeq B_i u^{\hat{M}_i-1}\;, \quad \bQ^i \simeq B^i u^{-\hat{M}_i}\;,
\eeq
where $\tilde{M}_a$, $a=1,\ldots,4$ and $\hat{M}_i$, $i=1,\ldots,4$ are expressed in terms of Cartan charges of ${\rm PSU}(2,2|4)$ (i.e. quantum numbers of the states)
\beqa\label{global_charges_all}
\tilde{M}_a&=&\left\{\frac{J_1+J_2-J_3}{2}+1, \frac{J_1-J_2+J_3}{2}, \frac{-J_1+J_2+J_3}{2}+1,\frac{-J_1-J_2-J_3}{2}\right\}, \notag \\
\hat{M}_i&=&\left\{\frac{1}{2}\left(\Delta-S_{+}+2\right), \frac{1}{2}\left(\Delta+S_{+}\right), \frac{1}{2}\left(-\Delta-S_{-}+2\right), \frac{1}{2}\left(-\Delta+S_{-}\right) \right\}\;,
\eeqa
where $S_{\pm}=S_1 \pm S_2$.

\begin{figure}
\centerline{
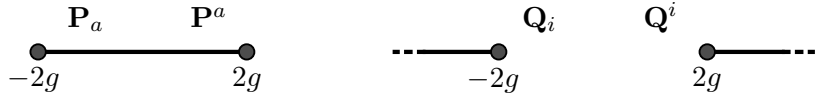}
\caption{Analytic structure of the $\bP$- and $\bQ$-functions on their defining sheet.}
\label{PQ_def_sheet}
\end{figure}

As the Q-system is generated by the set of $8$ basic Q-functions, we first ascribe them the analytic structure dictated by the classical limit of the Q-functions \cite{Gromov:2017blm}. The minimal choice of the cut structure consistent with the asymptotics \eqref{PQ_asymptotics} is presented on the Figure \ref{PQ_def_sheet}.
Then, according to the \eqref{QQ_relations} we can generate two versions of the Q-system: upper half plane and lower half plane analytic (UHPA and LHPA respectively). In what follows we will define by $\cQ_{a|i}$ the UHPA solution of the one of the QQ-relations from \eqref{QQ_relations}
\beq\label{eq_Qai}
\cQ_{a|i}^+-\cQ_{a|i}^-=\bP_a \bQ_i
\eeq
with the large $u$ asymptotic
\beq
\cQ_{a|i} \simeq -i\frac{A_a B_i}{-\tM_a+\hM_i}u^{-\tM_a+\hM_i}\;.
\eeq
Substitution of the formulas $\bQ_i=-\cQ_{a|i}^+ \bP^a$ or $\bP_a=-\cQ_{a|i} \bQ^i$ into \eqref{eq_Qai} allows to fix the products of the coefficients of the leading asymptotics of the $\bP$- and $\bQ$-functions and for $a_0,i_0=1,\ldots,4$ we get
\beq\label{AA_BB_general}
A_{a_0} A^{a_0}=i\frac{\prod\limits_{j=1}^4 \left(\tilde{M}_{a_0}-\hat{M}_j\right)}{\prod\limits_{\substack{b=1 \\ b \neq a_0}} \left(\tilde{M}_{a_0}-\tilde{M}_b\right)}\;, \quad B_{i_0} B^{i_0}=-i\frac{\prod\limits_{a=1}^4 \left(\hat{M}_{i_0}-\tilde{M}_a\right)}{\prod\limits_{\substack{j=1 \\ j \neq i_0}}^4 \left(\hat{M}_{i_0}-\hat{M}_j\right)}\;, 
\eeq
where there is no summation over the repeated indices $a_0$ and $i_0$.


An important consequence of the formula $\bQ_i=-\cQ_{a|i}\bP^a$ combined with the formula \eqref{eq_Qai} is the existence of a 4th order Baxter equation for the functions $\bQ_i$, $i=1,\ldots,4$ (see \cite{Alfimov:2014bwa} for the derivation)
\beqa\label{Baxter_4th_order}
\bQ_i^{[+4]}&-&\bQ_i^{[+2]}\left[D_1-\bP_a^{[+2]}\bP^{a[+4]}D_0 \right]+\bQ_i\left[D_2-\bQ_a \bP^{a[+2]}D_1+\bP_a \bP^{a[+4]}D_0 \right]- \notag \\
&-&\bQ_i^{[-2]}\left[\bar{D}_1+\bP_a^{[-2]}\bP^{a[-4]}\bar{D}_0 \right]+\bQ_i^{[-4]}=0\;,
\eeqa
where the functions $D_0$, $D_1$ and $D_2$ are given by
\beqa
D_0&=&\left|\begin{array}{llll}
\bP^{1[+2]} & \bP^{2[+2]} & \bP^{3[+2]} & \bP^{4[+2]} \\
\bP^{1} & \bP^{2} & \bP^{3} & \bP^{4} \\
\bP^{1[-2]} & \bP^{2[-2]} & \bP^{3[-2]} & \bP^{4[-2]} \\
\bP^{1[-4]} & \bP^{2[-4]} & \bP^{3[-4]} & \bP^{4[-4]}
\end{array}\right|\;, \quad
D_1=\left|(\begin{array}{llll}
\bP^{1[+4]} & \bP^{2[+4]} & \bP^{3[+4]} & \bP^{4[+4]} \\
\bP^{1} & \bP^{2} & \bP^{3} & \bP^{4} \\
\bP^{1[-2]} & \bP^{2[-2]} & \bP^{3[-2]} & \bP^{4[-2]} \\
\bP^{1[-4]} & \bP^{2[-4]} & \bP^{3[-4]} & \bP^{4[-4]}
\end{array}\right|, \notag \\
D_2&=&\left|\begin{array}{llll}
\bP^{1[+4]} & \bP^{2[+4]} & \bP^{3[+4]} & \bP^{4[+4]} \\
\bP^{1[+2]} & \bP^{2[+2]} & \bP^{3[+2]} & \bP^{4[+2]} \\
\bP^{1[-2]} & \bP^{2[-2]} & \bP^{3[-2]} & \bP^{4[-2]} \\
\bP^{1[-4]} & \bP^{2[-4]} & \bP^{3[-4]} & \bP^{4[-4]}
\end{array}\right|\;,
\eeqa
while the bars over $\bar{D}_1$ and $\bar{D}_2$ are understood as the complex conjugation $\bar{f}(u)=\overline{f(\bar{u})}$ of the functions defined above. The same equation for $\bQ^i$, $i=1,\ldots,4$ is valid, if we turn the functions $\bP^a$ into $\bP_a$ for $a=1,\ldots,4$.

One can see from \eqref{Baxter_4th_order} that if we are far away (high or low enough) from the real axis in the complex plane, the cut structures (see the Figure \ref{PQ_def_sheet}) of $\bP$- and $\bQ$-functions do not contradict each other. But as we approach the real axis, this is no longer the case and we have to cross the cut.
This makes the QQ-relations ambiguous in the vicinity of the cuts, as the shifts by $\pm i/2$
may cross the cut and one may ask which contour to use to reach $u\pm i/2$ from the point $u$ or, in other words, which branch of the multi-valued function to use.

The way to resolve the analytic continuation ambiguity is to interpret the values of the $\bQ$-function in the upper half plane, far enough from the cuts, as a $\bQ$-function with the upper index $\bQ^i$ and in the lower half plane the same $\bQ$-function should
obey the QQ-relations as if it was a function with lower index $\bQ_i^{\uparrow}$.
I.e. in the lower half plane this function satisfies \eqref{Baxter_4th_order}, whereas in the upper half plane it satisfies the same equation with $\bP_a$ interchanged with $\bP^a$. 
The ${\uparrow}$ is added to indicate that $\bQ_i^{{\uparrow}}$ does not have cuts in the lower half plane.
To close the system of equations we notice that there is yet another way to build the set of functions $\bQ_i^{{\uparrow}}$, satisfying \eqref{Baxter_4th_order} and having no cuts in the lower half plane. One can, starting from $\bQ^i$, which has no cuts in the upper half plane
and $\bP^a$ -- and build $\bQ_i^{\downarrow}$ using QQ-relations. Then the complex conjugate of
$\bQ_i^{\downarrow}$ will satisfy \eqref{Baxter_4th_order} and will have no cuts in the lower half plane.

Indeed, due to the property, valid  for real $S_1$, $S_2$ and $\Delta$ we can always assume that
\beq\label{P_conjugation}
\bar{\bP}_a=C_a^b \bP_b\;, \quad \bar{\bP}^a=-C^a_b \bP^b\;, \quad C=\textrm{diag}\{1,1,-1,-1\}
\eeq
due to the H-symmetry. Thus if we complex conjugate the Baxter equation \eqref{Baxter_4th_order}, we get the same equation for $\bar{\bQ}_j$. 
This implies that $\bQ_i^{\downarrow}(u)$
is a linear combination of $\bar{\bQ}_j$ with regular coefficients
\beq\label{QQbar_transform}
\bQ^i(u)=M^{ij}(u) \bar{\bQ}_j(u)\;, \quad \bQ_i(u)=\left(M^{-t}\right)_{ij}(u) \bar{\bQ}^j(u)\;,
\eeq
where $-t$ denotes the inversion and transposition of the matrix.
Technically it will be more convenient to 
work with short cuts $[-2g,2g]$. We can connect the branch points of $\bQ$ the way we like, but this would create an infinite ladder of cuts
in the lower half plane. It will also modify the notion of the conjugate function, as the conjugation now will also involve the analytic continuation under the cut.
So we conclude that in the short cut conventions
the gluing condition reads as
\beq\la{gluingeq}
\tilde{\bQ}^i(u)=M^{ij}(u) \bar{\bQ}_j(u)\;, \quad \tilde{\bQ}_i(u)=\left(M^{-t}\right)_{ij}(u) \bar{\bQ}^j(u)\;,
\eeq
which we will use further.
Let us now list the properties of $M^{ij}(u)$, which is called the gluing matrix. The details of the derivation of these properties can be found in \cite{Alfimov:2018cms}. They are:
\begin{itemize}

\item $M^{ij}(u)$ is an $i$-periodic matrix of H-transformation.

\item $M^{ij}(u)$ is analytic in the whole complex plane.

\item $M^{ij}(u)$ is hermitian as a function
\beq
\bar{M}^{ij}(u)=M^{ji}(u)\;.
\eeq

\end{itemize}



There exist another additional symmetry of the $\bP$-functions. The $\bP$-functions of the states with the Cartan charges
\beqa\label{global_charges}
\tilde{M}_a&=&\left\{2, 1, 0, -1 \right\}, \\
\hat{M}_i&=&\left\{\frac{1}{2}\left(\Delta-S_{+}+2\right)\;, \frac{1}{2}\left(\Delta+S_{+}\right)\;, \frac{1}{2}\left(-\Delta-S_{-}+2\right)\;, \frac{1}{2}\left(-\Delta+S_{-}\right) \right\}\;, \notag
\eeqa
where $S_{\pm} \equiv S_1 \pm S_2$, have the certain parity
\beq\label{P_parity}
\bP_a(-u)=(-1)^{a+1}\bP_a(u)\;, \quad \bP^a(-u)=(-1)^a \bP^a(u)\;.
\eeq
The conjugation symmetry and the symmetry above \eqref{P_parity} lead to two additional constraints \cite{Alfimov:2018cms} on the gluing matrix. Let us first concentrate on the physical states, when $S_1$ and $S_2$ are integer and have the same parity\footnote{This is dictated by the cyclicity condition for the states in the $\mathfrak{sl}(2)$ Heisenberg spin chain.}. As it follows from the power-like large $u$ asymptotics of the $\bQ$-functions in this case, the only possible ansatz for the gluing matrix is a constant matrix. In \cite{Gromov:2014caa,Alfimov:2018cms} it was shown that for the physical length-2 states with the charges \eqref{global_charges} from the abovementioned properties of the gluing matrix and two additional constraints mentioned earlier it follows that 
\beq\label{int_ch_cond2}
M^{ij}\left(e^{i\pi\left(\hat{M}_i-\hat{M}_j\right)}+1\right)=0\;.
\eeq

From \eqref{int_ch_cond2}, we immediately see that only when the difference between the charges is an odd integer, $M^{ij}$ is non-zero. It is the case only for
\beqa\label{M_int_ch_nonzero}
\hat{M}_1-\hat{M}_2&=&-S_1-S_2+1\;, \\
\hat{M}_3-\hat{M}_4&=&-S_1+S_2+1\;. \notag
\eeqa
Therefore, taking into account the hermiticity of the gluing matrix, for integer $S_1$ and $S_2$, that have the same parity\footnote{At one loop this is dictated by the cyclicity condition for the states in the $\mathfrak{sl}(2)$ Heisenberg spin chain, appearing in the perturbation theory.} (i.e. for the physical states) the equations \eqref{int_ch_cond2} lead to the gluing matrix
\beq\label{gluing_matrix_spins_same_parity}
M^{ij}=\left(
\begin{array}{cccc}
0 & M^{12} & 0 & 0 \\
\bar{M}^{12} & 0 & 0 & 0 \\
0 & 0 & 0 & M^{34} \\
0 & 0 & \bar{M}^{34} & 0
\end{array}
\right)\;.
\eeq

In the case of at least one of the spins $S_1$ and $S_2$ being non-integer, we see, that because all differences $\hat{M}_i-\hat{M}_j$ are non-integer the equations \eqref{int_ch_cond2} can have only zero matrix as a solution, if this matrix is assumed to be constant. This leads us to the conclusion that the gluing matrix cannot be constant anymore for non-integer spins. The minimal way to do this keeping it $i$-periodic would be to add exponential contributions and get the following gluing matrix
\beqa\label{gluing_conditions_non_integer_spins_solution}
M&=&\left(\begin{array}{cccc}
M_1^{11} & M_1^{12} & M_1^{13} & M_1^{14} \\
\bar{M}_1^{12} & 0 & 0 & 0 \\
\bar{M}_1^{13} & 0 & M_1^{33} & M_1^{34} \\
\bar{M}_1^{14} & 0 & \bar{M}_1^{34} & M_1^{44} \\
\end{array}\right)+ \notag \\
&+&\left(\begin{array}{cccc}
0 & 0 & M_2^{13} & M_2^{14} \\
0 & 0 & 0 & 0 \\
\bar{M}_2^{13} & 0 & 0 & 0 \\
\bar{M}_2^{14} & 0 & 0 & 0 \\
\end{array}\right)e^{2\pi u}
+\left(\begin{array}{cccc}
0 & 0 & M_3^{13} & M_3^{14} \\
0 & 0 & 0 & 0 \\
\bar{M}_3^{13} & 0 & 0 & 0 \\
\bar{M}_3^{14} & 0 & 0 & 0 \\
\end{array}\right)e^{-2\pi u}\;.
\eeqa

To summarize, the key observation here is that if we want to consider non-physical states with $S_1$ or $S_2$ being non-integer, this inevitably requires modification of the gluing conditions. From now on let us use the notations for the spins coming from BFKL physics $S_1=S$ and $S_2=n$.

In the next Subsection we will briefly describe how the QQ-relations together with the gluing condition \eq{gluingeq} can be used to compute numerically the function $S(\Delta,n)$.

\subsection{Numerical solution}\label{LO_BFKL_num}

The QQ-system together with the gluing conditions allows for the efficient numerical algorithm, developed in \cite{Gromov:2015wca}. This method was described in details with the {\it Mathematica} code attached in a recent review \cite{Gromov:2017blm}. Here we describe the main steps very briefly.

First, one notices that the $\bP_a$ and $\bP^a$ functions, $a=1,\ldots,4$, which have only one short cut $[-2g,2g]$ on the main sheet of the Riemann surface, can be parametrized very efficiently as rapidly converging series expansions in powers of $1/x(u)$, where $x(u)=(u+\sqrt{u-2g}\sqrt{u+2g})/(2g)$. After that one can recover the whole Q-system in terms of these expansion coefficients. For that one first
uses 
\beq\label{eq_Qai2}
\cQ_{a|i}^+-\cQ_{a|i}^-=-{\cQ}_{b|i}^+ \bP_a {\bP}^b\;,
\eeq
which follows directly from \eq{eq_Qai} and $\bQ_i=-\cQ_{a|i}\bP^i$, to solve for $\cQ_{a|i}$. 
After that one finds $\bQ_i$ and $\tilde\bQ_i$ from
\beq
{\bQ}_i=-{\cQ}_{a|i}^+ {\bP}^a\;\;,\;\;
\tilde {\bQ}_i=-{\cQ}_{a|i}^+ \tilde {\bP}^a\;.
\eeq
Note that this involves $\tilde {\bP}^a$, which is the same series as ${\bP}^a$, but with $x\to 1/x$ for $u\in[-2g,2g]$. In this way we reconstruct both $\bQ_i$ and $\tilde\bQ_i$ in terms of the expansion coefficients in $\bP_a$ and $\bP^a$.
Finally, one fixes these coefficients from the gluing condition \eq{gluingeq}.

In practice for the numerical purposes one truncates the series expansion in $\bP$. In this case the gluing condition cannot be satisfied exactly. 
The strategy is to minimize the discrepancy in the gluing condition by adjusting the coefficients numerically with a variation of a Newton method.

Application of the numerical procedure allows to calculate the dependence of the intercept function $j=S(\Delta=0,n=0)+2$ for zero conformal spin \cite{Gromov:2014bva} on the `t Hooft coupling $\lambda$, which is drawn on the Figure \ref{fig:intercept}. By fitting the polynomial dependence of this quantity on the coupling, we obtain several first coefficients
\beqa\la{intercept_num_fit}
j(\lambda)&=&1+0.07023049277268284\; \lambda-0.00337167607361\; \lambda
   ^2\\ \nonumber
   &-&0.00064579607573\; \lambda ^3
   + 0.0002512619258\; \lambda ^4+\mathcal{O}(\lambda^5)\;.
\eeqa
\begin{figure}
\centerline{
\includegraphics[scale=1]{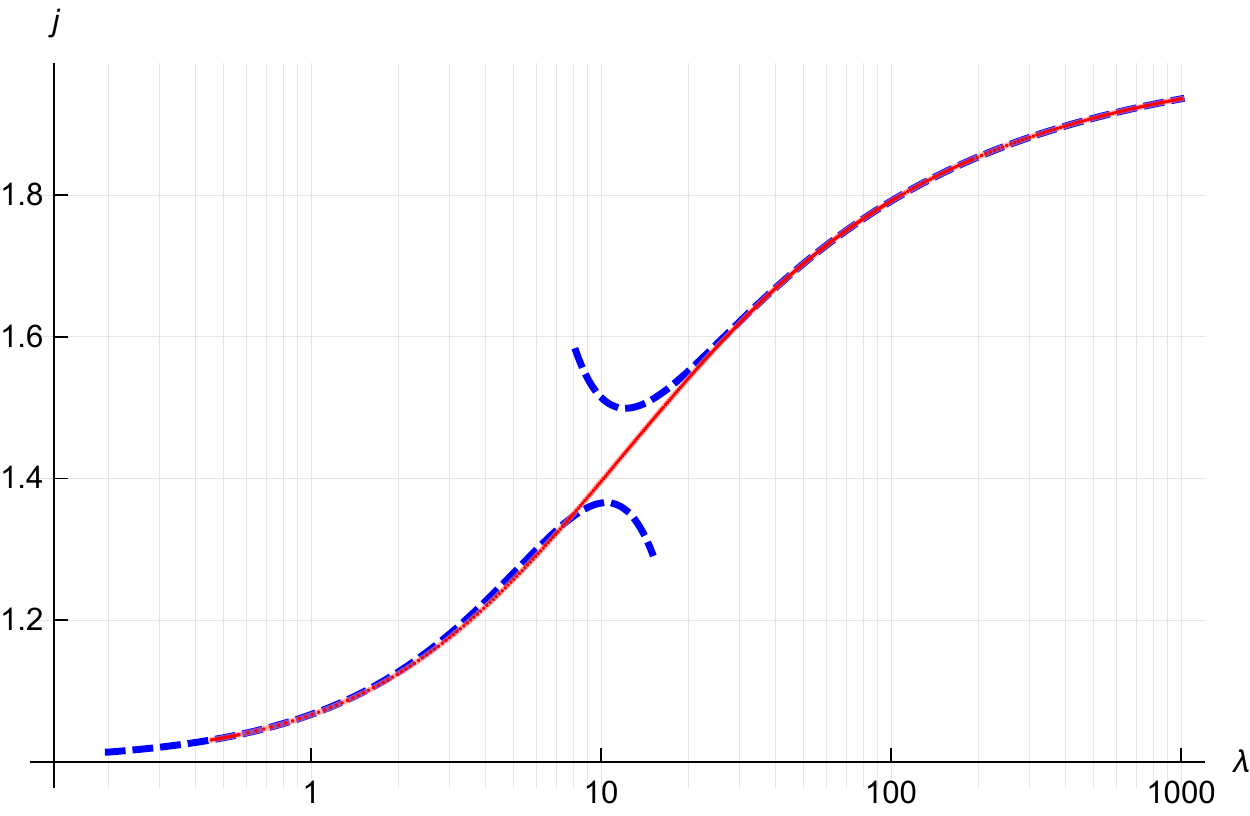}}
\caption{The dependence of the intercept function $j$ on the `t Hooft coupling $\lambda$ in the logarithmic scale. Data obtained from the numerical procedure are depicted in red. It represents a perfect interpolation between the weak \cite{Kotikov:2002ab,Brower:2006ea} and strong \cite{Costa:2012cb,Kotikov:2013xu,Brower:2013jga,Gromov:2014bva} coupling results (blue dashes).}
\la{fig:intercept}
\end{figure}
Comparing the first two coefficients of \eqref{intercept_num_fit}, we find complete agreement with the LO \eqref{lon} and NLO BFKL \eqref{NLOoriginal_SYM} kernel eigenvalues calculated at $\Delta=0$ and $n=0$\footnote{After renormalizing these results according to the change of the expansions parameter from $g^2$ to $\lambda$.}.

Moreover, we can use our numerical algorithm to calculate not only the intercept functions, but also the BFKL kernel eigenvalues at different orders in $g$. Namely, in \cite{Gromov:2015vua} from fitting the numerical values of the spin $S(\Delta,n)$ for $\Delta=0.45$ and $n=0$ for different values of the coupling constant one was able to extract the numerical values of the coefficients $\chi_{\textrm{N}^k \textrm{LO}}(\Delta=0.45)$ from \eqref{S_exp_BFKL}, which are listed in the Table~\ref{num_values_BFKL}.

\begin{table}[h!]
\begin{center}
\small{\caption{Numerical values of the BFKL kernel eigenvalue at different orders.}
\label{num_values_BFKL}
{\begin{tabular}{ccc}
\hline
Order & Value & Error \\
\hline
${\rm N^2LO}$
& $10774.635818847176637957593127192456995929170948057653783424533229$
& $10^{-61}$
\\
${\rm N^3LO}$ 
& $-366393.2052053917038937903507478544549935531959333919163403836$
& $10^{-56}$
\\
${\rm N^4LO}$
& $1.332736355681126915694044310369828561521940588979476878854 \times 10^7$
& $10^{-51}$
\\
${\rm N^5LO}$
& $-4.921740136657916500913955552075070060721450958436559876 \times 10^8$
& $10^{-47}$ \\
\hline
\end{tabular}}}
\end{center}
\end{table}
The first line of the Table \ref{num_values_BFKL} contains the numerical value of the NNLO BFKL kernel eigenvalue for $\Delta=0.45$, which was computed analytically for the first time 
in \cite{Gromov:2015vua} by the QSC method. In that work there was shown that the calculated numerical value in the Table \ref{num_values_BFKL} coincides with the value of the analytic result for $\Delta=0.45$ with numerical accuracy $10^{-61}$.

\begin{figure}
\centerline{
\includegraphics[width=0.7\linewidth]{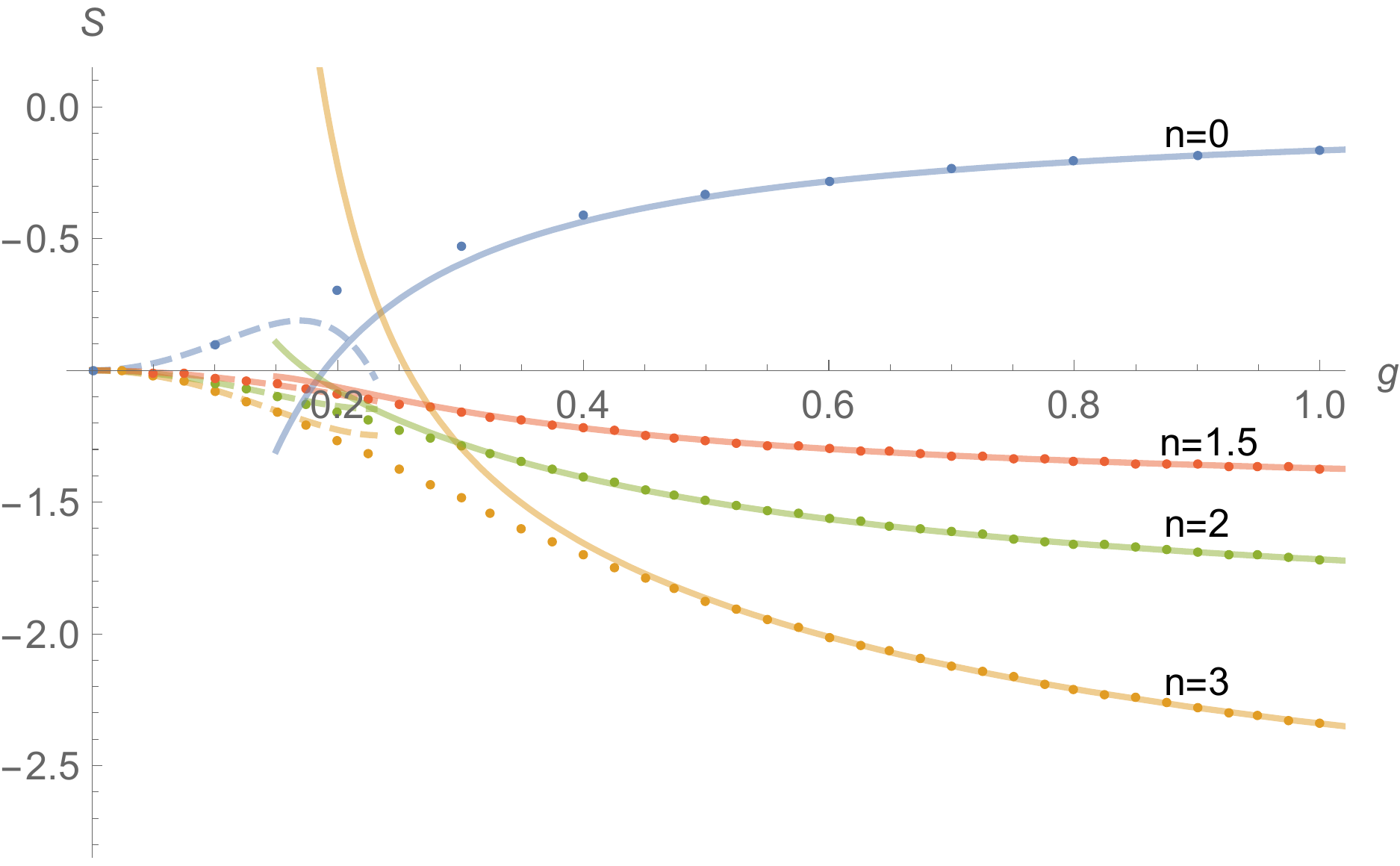}}
\caption{The dependence of the numerical values of the spin $S(0,n)$ on the coupling constant $g$, calculated for the conformal spins $n=0$, $n=3/2$, $n=2$ and $n=3$ (dots). The dashed lines represent weak coupling expansion and the continuous lines represent the strong coupling expansion of the spin obtained in \cite{Alfimov:2018cms}.}
\label{intercept_numerics}
\end{figure}

In addition, the numerical algorithm can be used to explore the spectrum for non-zero values of conformal spin $n$ (to see the algorithm for different values of $\Delta$, $n$ and $g$ at work one can use the {\it Mathematica} file \verb"code_for_arxiv.nb" from the \cite{Alfimov:2018cms} arXiv submission). One can find the plots of the spin $S(0,n)$, which differs from the intercept function $j(n)$ by the additive constant $2$, for different (even non-integer) values of the conformal spin $n$ on the Figure \ref{intercept_numerics}.

To conclude, the numerical method of \cite{Gromov:2015wca} allowed to obtain the BFKL kernel eigenvalue non-perturbatively with huge precision.

A number of analytic methods was developed to solve the QSC at small $g$ \cite{Marboe:2014gma,Marboe:2014sya,Gromov:2015vua,Marboe:2017dmb,Marboe:2018ugv}. Unfortunately, those methods rely on a particular basis of $\eta$-functions, which works very well for the local operators of for BFKL eigenvalues at some integer values of $\Delta$, but are not sufficient in general. In the next section we will use an alternative method to obtain the LO BFKL kernel eigenvalue analytically. Recently, new very promising methods were developed in \cite{Lee:2017mhh,Lee:2019bgq,Lee:2019oml}, based on the Mellin transformations, which could allow for a systematic analytic calculation of $S(\Delta)$ order by order in $g^2$ for generic $\Delta$.

\subsection{Analytic results from QSC}\label{LO_BFKL}

Apart from the numerical results mentioned in the previous Section, a number of analytic results related to the BFKL spectrum was obtained recently using the QSC methods in~\cite{Gromov:2014bva,Alfimov:2014bwa,Gromov:2015vua,Gromov:2015dfa,Gromov:2016rrp,Alfimov:2018cms}. To mention a few:
\begin{itemize}

\item NNLO BFKL kernel eigenvalue for the conformal spin $n=0$.

\item Intercept function for arbitrary conformal spin $n$ up to NNLO order and partial result at NNNLO order.

\item Strong coupling expansion of the intercept function for arbitrary value of the conformal spin $n$.

\item Slope-to-slope function in the BPS point $\Delta=2$, $S=0$ and $n=0$ at all loops.
    
\item Slope-to-intercept $dS(\Delta,n)/dn$ and curvature $d^2 S(\Delta,n)/d\Delta^2$ functions in the BPS point $\Delta=0$, $S=-1$ and $n=1$ at all loops.
\end{itemize}

Here we demonstrate the power of the QSC method deriving the leading order Faddeev-Korchemsky Baxter equation with non-zero conformal spin for Lipatov spin chain analytically. 

As explained above, we need to study the regime of the anomalous dimensions of the twist-2 operators, when the coupling constant $g^2 \rightarrow 0$, such that $S\equiv S_1(\Delta,g)$ becomes $-1+{\cal O}(g^2)$. According to the discussion in the previous section this is the case for $|\Delta|<1+|n|$.
In other words one can say that we keep the ratio $\Lambda\equiv g^2/(S+1)$ finite,
whereas the combination $w\equiv S+1={\cal O}(g^2)$ can be used as a small expansion parameter.
The second spin $S_2=n$ plays the role of a parameter. 

In order to reproduce the Faddeev-Korchemsky Baxter equation we are going to utilise the subset of the QQ-relations known as $\bP\mu$-system. The $\bP\mu$-system is represented by the functions $\bP_a(u)$, $\bP^a(u)$, which we introduced before and of an anti-symmetric matrix $\mu_{ab}(u)$ (see \cite{Gromov:2013pga,Gromov:2014caa} for the detailed description). They satisfy the following equations
\beqa\label{Pmu_equations}
&\tilde{\mu}_{ab}-\mu_{ab}=\bP_a \tilde{\bP}_b-\bP_b \tilde{\bP}_a\;, \quad &\tilde{\bP}_a=\mu_{ab}\bP^b\;, \\
&\tilde{\mu}^{ab}-\mu^{ab}=\bP^a \tilde{\bP}^b-\bP^b \tilde{\bP}^a\;, \quad &\tilde{\bP}^a=\mu^{ab}\bP_b\;, \notag \\
&\bP_a \bP^a=0\;, \quad \mu_{ab}\mu^{bc}=\delta_a^c\;, \quad &\tilde{\mu}_{ab}=\mu_{ab}^{++}\;, \quad \tilde{\mu}^{ab}=\mu^{ab++}\;. \notag
\eeqa
Before proceeding we remind a couple of notations. Our notation for the BFKL scaling parameter is $w=S+1$. It is also convenient to use the notation $\Lambda=g^2/w$.

To start solving the $\bP\mu$-system in the BFKL regime we have to determine the scaling of the $\bP$-, $\bQ$- and $\mu$-functions in the small $w$ limit. In what follows we are going to use the arguments from \cite{Alfimov:2014bwa}, thus as from \eqref{AA_BB_general} for the length-2 state in question \eqref{length_2_operator} in the BFKL limit $A_a A^a=\cO(w^0)$ for $a=1,\ldots,4$ and $B_i B^i=\cO(w^0)$, these functions can be chosen to scale as $w^0$
\beqa\label{P_BFKL_scaling}
\bP_a=\bP^{(0)}_a+\mathcal{O}(w)\;,& \quad &\bP^a=\bP^{(0)a}+\mathcal{O}(w) \\
\bQ_i=\bQ^{(0)}_i+\mathcal{O}(w)\;,& \quad &\bQ^i=\bQ^{(0)i}+\mathcal{O}(w)
\eeqa
and the $\mu$-functions scale as $w^{-2}$
\beq
\mu_{ab}=w^{-2}\left(\mu^{(0)}_{ab}+\mathcal{O}(w)\right)\;, \quad \mu^{ab}=w^{-2}\left(\mu^{(0)ab}+\mathcal{O}(w)\right)\;.
\eeq
Additionally, as all the $\bP$-functions for the length-2 states being considered possess the certain parity from the $\bP\mu$-system equations \eqref{Pmu_equations} we can conclude that the functions $\mu_{ab}^+(u)$ have the certain parity.

Let us restrict ourselves from now on in this Section to the case of integer conformal spin $S_2=n$. We know that the $\bP$-functions have only one cut on one of the sheets, therefore they can be written as a series in the Zhukovsky variable $x(u)=(u+\sqrt{u^2-4\Lambda w})/(2\sqrt{\Lambda w})$\footnote{Here we rewrote the Zhukovsky variable using $g^2=\Lambda w$.}. Then, we are also allowed to apply the certain H-transformation to the $\bP$-functions, which do not alter their asymptotics and parity. Applying this transformation, we can set the coefficients $A_1=A_2=-A^3=A^4=1$ and some other coeficients in the series in $x(u)$ to $0$. Thus, we arrive to the formulas in the LO
\beqa\label{P_BFKL_LO}
\bP^{(0)}_1=\frac{1}{u^2}\;, \quad \bP^{(0)}_2=\frac{1}{u}\;,& \quad &\bP^{(0)}_3=A_3^{(0)}\;, \quad \bP^{(0)}_4=A_4^{(0)}u+\frac{c_{4,1}^{(1)}}{\Lambda u}\;, \\
\bP^{(0)1}=A^{1(0)}u+\frac{c^{1,1(1)}}{\Lambda u}\;,& \quad &\bP^{(0)2}=A^{2(0)}\;, \quad \bP^{(0)3}=-\frac{1}{u}\;, \quad \bP^{(0)4}=\frac{1}{u^2}\;, \notag
\eeqa
where
\beqa\label{A3A4_BFKL_LO}
A_3^{(0)}&=&\frac{(\left(\Delta-n\right)^2-1)(\left(\Delta+n\right)^2-9)}{32i}\;, \\
A_4^{(0)}&=&\frac{(\left(\Delta+n\right)^2-1)(\left(\Delta-n\right)^2-25)}{96i} \notag
\eeqa
originate from the expansion of \eqref{AA_BB_general} at small $w$ and $c_{4,1}^{(1)}$ and $c^{1,1(1)}$ are some yet unknown coefficients.

The situation with the asymptotics of the $\mu$-functions is more subtle. For non-zero $S$ it appears that the asymptotics of the $\mu$ cannot be power-like anymore \cite{Gromov:2014eha} and the minimal modification of the leading asymptotics of the $\mu$-functions with the lower indices at $u \rightarrow \pm\infty$ is
\beq\label{mu_asymptotics}
\mu_{ab} \sim (u^{-S-1},u^{-S},u^{-S+1},u^{-S+1},u^{-S+2},u^{-S+3})e^{2\pi |u|}\;,
\eeq
while the $\mu$-functions with the upper indices have the same asymptotics but in the reverse order. The fact that the asymptotics should be modified as \eq{mu_asymptotics} can be derived from the fact that the gluing matrix~\eq{gluing_conditions_non_integer_spins_solution} asymptotic receives additional exponential terms. 
In the work \cite{Alfimov:2018cms} it was shown, that the correct ansatz for the $\mu$-functions in the LO would be polynomials of the powers consistent with the asymptotics \eqref{mu_asymptotics} for $S=-1+w$ and $w$ close to $0$. We do not write this ansatz explicitly as it is not important for further discussion.




Substituting \eqref{P_BFKL_LO} and the ansatz for the $\mu$-functions into \eqref{Pmu_equations} in the LO and from the analyticity properties of the $\bP$- and $\mu$-functions we fix the constants $c^{1,1(1)}$ and $c_{4,1}^{(1)}$
\beq\label{coefficients_analyticity_constraints}
c^{1,1(1)}=-c_{4,1}^{(1)}\;, \quad c_{4,1}^{(1)}=-\frac{i\Lambda}{96}\left(\left(\Delta^2-1\right)^2-2\left(\Delta^2+1\right)n^2+n^4\right)
\eeq
and the $\mu$-functions in the LO up to an overall constant (we do not write them explicitly as they are not relevant for further discussion).



After the substitution of the LO $\bP$-functions \eqref{P_BFKL_LO} together with \eqref{coefficients_analyticity_constraints} into \eqref{Baxter_4th_order} in the LO (for the same result with the zero conformal spin see \cite{Alfimov:2014bwa}), one can derive, that this 4th order Baxter equation factorizes
\beqa\label{Baxter_4th_order_LO}
\left[(u+2i)^2 D+(u-2i)^2 D^{-1}\right.&-&\left.2u^2-\frac{17-(\Delta+n)^2}{4}\right] \times \\
&\times& \left[D+D^{-1}-2-\frac{1-(\Delta-n)^2}{4u^2}\right]\bQ^{(0)}_j=0\;, \notag
\eeqa
where $D=e^{i\partial_u}$ is the shift operator (the same equation with $n$ replaced by $-n$ is true for $\bQ^{(0)j}$, $j=1,\ldots,4$). Thus, two out of four linearly independent solutions of the $4^{\rm th}$ order Baxter equation can be found by soliving the second order Baxter equation
\beq\label{lQ13_Baxter_LO}
\bQ^{(0)++}+\bQ^{(0)--}+\left(-2+\frac{(\Delta-n)^2-1}{4u^2}\right)\bQ^{(0)}=0\;.
\eeq

Redefining the Q-function to be $Q=\frac{\bQ_j}{u^2}$ we exactly reproduce the Baxter equation for the $SL(2,\mathbb{C})$ spin chain \cite{Faddeev:1994zg,DeVega:2001pu,Derkachov:2001yn,Derkachov:2002pb,Derkachov:2002wz} from the QSC for \sym\, which is a highly non-trivial test of the all-loop 
integrability of this theory. This derivation established the desired connection between the integrability in the 4D gauge theory in the BFKL limit and integrability of the \sym. Having this result, one can go further and by using QSC explore many more quantities, such as NNLO BFKL kernel eigenvalue, numerical twist-2 and length-2 operator trajectories, intercept function, slope, curvature and slope-to-intercept function etc. (see \cite{Gromov:2015wca,Gromov:2015vua,Alfimov:2018cms} and references therein).

\section{Fishnet and BFKL}
Above we described how 
the integrability, discovered in the BFKL regime of the QCD, can be understood as a part of a more general integrable structure of ${\cal N}=4$ SYM.
In fact the data coming from the BFKL regime has played an essential role in fixing the dressing phase of the Beisert-Staudascher equations, which eventually resulted in our current understanding of integrability in this model. 

Another extremely popular recent topic --  Sachdev-Ye-Kitaev (SYK) model also has many similarities at the technical level with the problem of BFKL kernel diagonalization~ \cite{Gross:2017hcz,Klebanov:2016xxf,Murugan:2017eto,Kitaev} (see also a recent review dedicated to this topic~\cite{Rosenhaus:2019mfr}).

In this section, we will describe yet another example of CFT tightly related to the LO BFKL physics -- the Fishnet CFT at any dimension~(FCFT$_D$). We will explain its Feynman graph content (in planar approximation big graphs have the shape of regular square lattice -- at the origin of the name ``fishnet") and reveal the origins of its integrbility, described in terms of the conformal $SO(1,D+1)$ spin chain.
We will demonstrate how, in particular case of $D=2$ and zero conformal spin, the problem of computing the anomalous dimensions of the simplest operators, described by so-called wheel graphs,  boils down to the calculation of the spectrum of Lipatov's multi-reggeon  Hamiltonian.

\subsection{Fishnet integrable model}

The $D-$dimensional fishnet biscalar conformal field theory (dubbed usually as fishnet CFT, or FCFT$_D$) is defined by the  Lagrangian~\cite{Kazakov:2018qez}
\begin{equation}\label{bi-scalarL-D}
    {\cal L}_{d}=  N_c\,\tr
    [\bar X \,\, (-\p_\mu \p^\mu)^{\frac{D}{4}-\omega}\,X + \bar Z \,\, (-\p_\mu \p^\mu)^{\frac{D}{4}+\omega}\,Z 
+(4 \pi)^{\frac{D}{2}} \xi^2 \bar X \bar Z X Z]\;,
  \end{equation} 
where $X,Z$ are complex $N_c\times N_c$ matrix fields and $\bar X\equiv X^\dagger,\bar Z\equiv Z^\dagger$ are their Hermitian conjugates. The differential operator in an arbitrary power is defined in a standard way, as an integral operator.  The action \eqref{bi-scalarL-D} should be supplemented with  certain double-trace counterterms described in~\cite{Fokken:2014soa,Sieg:2016vap,Grabner:2017pgm,Gromov:2018hut} (see also \cite{Kazakov:2018hrh} and references therein). 

The FCFT$_4$ case of the model (in $D=4$ dimensions), for the particular ``isotropic" case $\omega=0$, has a local Lagrangian. It  was proposed in~\cite{Gurdogan:2015csr} as a specific double scaling limit of the $\gamma$-deformed ${\cal N}=4$ SYM theory, combining strong (imaginary) deformation and weak coupling. The effective coupling constant is $\xi^2=g_{\rm YM}^2 N_c e^{-i\gamma_3}/(4\pi)^2$, where the `t Hooft coupling $\lambda=g_{YM}^2 N_c \rightarrow 0$ and the deformations parameter $\gamma_3 \rightarrow i\infty$. In this limit, all the fields except for two scalars get decoupled leading to \eqref{bi-scalarL-D} with $D=4$ and $\omega=0$, but the model retains the $SU(N_c)$ global symmetry, which is a remnant of the gauge symmetry of the original ${\cal N}=4$~SYM theory. In the planar limit $N_c\to\infty$, the FCFT is dominated by so-called fishnet Feynman graphs which have the structure of the regular square lattice. Such graphs appear to be integrable~\cite{Zamolodchikov:1980mb,Chicherin:2012yn,Gromov:2017cja}. 
For infinitely long operators at a critical coupling $\xi_*$ those graphs have a continuous limit which is believed to be described by a specific $O(2,4)$ 2D $\sigma$-model~\cite{Basso:2018agi}. 
At strong coupling $\xi\to\infty$
the ${\rm FQFT}_4$ has a classical description in terms of a dual string-bit {\it fish-chain} model as it was derived  in~\cite{Gromov:2019aku}.
This exact duality persists at the quantum level too~\cite{Gromov:2019jfh}.

The model has already a rich history of studying various physical quantities, such as  explicit computations of spectra of local operators~\cite{Gurdogan:2015csr,Caetano:2016ydc} and 4-point correlation functions~\cite{Basso:2017jwq,Derkachov:2018rot,Grabner:2017pgm,Gromov:2018hut}. The QSC formalism, adopted for the $\gamma$-deformed ${\cal N}=4$ SYM theory~\cite{Kazakov:2015efa}, appeared in this limit particularly efficient for computations of anomalous dimensions of operators of the type $\tr Z^L(\bar XX)^M+{\rm permutations}$
and later was extended to a much wider class of operators in~\cite{Gromov:2019jfh}. The FCFT$_4$ also appears to possess a rich moduli space of flat vacua, which are quantum-mechanically stable in spite of the absence of supersymmetry in the model~\cite{Karananas:2019fox} (in the planar limit).

\subsection{Spectrum of Fishnet model and LO BFKL}

At the technical level there is a number of places where the study of the spectrum of the fishnet models goes along  similar steps as the problem of the BFKL spectrum. 
The similarity of the integrable structure of both models is discussed below. Here we just consider a simple example of a dimension of operators of the type ${\rm tr}\Box^n Z^2X^2$ in $D=4,\;w=0$, where $\Box$ is the 4D Laplace operator.
The dimensions $\Delta$
of these operators can be obtained by solving the equation~\cite{Gromov:2018hut}
\beq\la{E2}
\frac{1}{\xi^4}=\frac{\psi ^{(1)}\left(\frac{1}{4}
        (4-\Delta )\right)-\psi
        ^{(1)}\left(\frac{1}{4} (6-\Delta
        )\right)-\psi
        ^{(1)}\left(\frac{\Delta
        }{4}\right)+\psi
        ^{(1)}\left(\frac{1}{4} (2+\Delta)\right)}{ (\Delta
        -2)}\;,
\eeq
which is reminiscent of the LO BFKL eigenvalue \eq{LO_BFKL_eigenvalue}, written in the form
\beq\label{pomeron}
\frac{1}{g^2}=\frac{4}{S+1} \left(-\psi\left(\frac{1+\Delta}{2}\right)-\psi\left(\frac{1-\Delta}{2}\right)+2\psi(1)\right)+\mathcal{O}\left(\frac{g^2}{S+1}\right)\;.
\eeq
The derivation of~\cite{E2}
from a Baxter equation is also very similar, to what is described in the Section \ref{LO_BFKL}. 

Whereas the above analogy is demonstrated only at the structural level, below we give a more precise relation between the BFKL Hamiltonian and the graph-building operator of the fishnet theory in $D=2$ and a  particular, spin zero representation for physical spins.

\subsection{Graph-building operator and BFKL kernel}

\begin{figure}[ht]
\begin{minipage}[h]{0.49\linewidth}
\center{\includegraphics[angle=-90,width=2.1in]{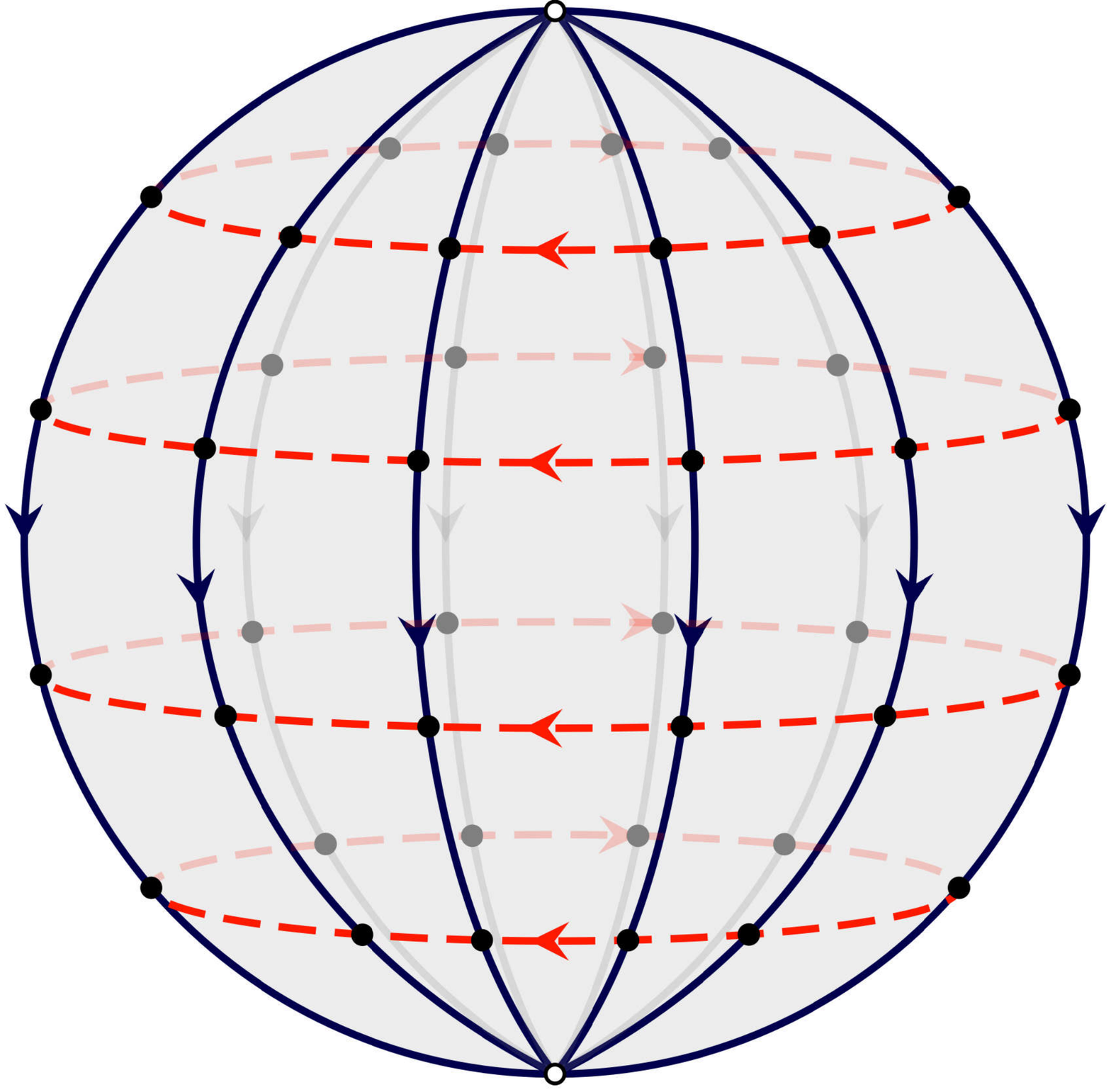}
\caption{Planar Fishnet graphs of ``globe" type arising in the computation of correlator \eqref{eq:corr}. The edges belonging to red ``parallels" and the black ``meridians" correspond to propagators for two different scalar fields of the Lagrangian~\eqref{bi-scalarL-D}. Due to the chiral properties of vertices of this Lagrangian, no other planar diagrams in each loop order exist.}
\label{globe_Fishnet}}
\end{minipage}
\hfill
\begin{minipage}[h]{0.49\linewidth}
\center{\includegraphics[angle=-90,width=2.1in]{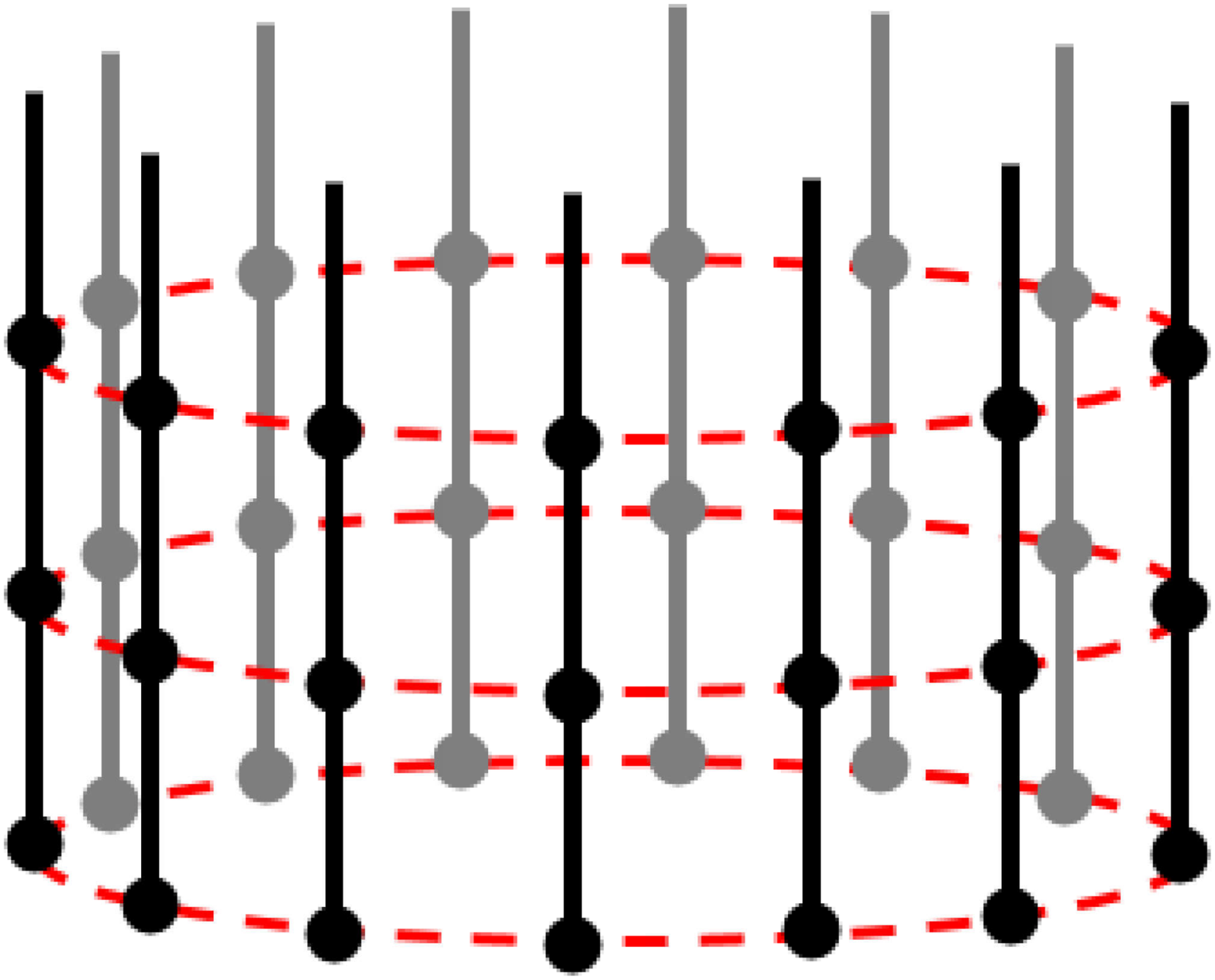}
\caption{The cyllinder fishnet configuration resulting from the opening of the ends of propagators at the south and north poles of the globe (i.e. separating their coordinates) and amputating the propagators at the south pole. Such graphs can be computed by the Bethe-Salpeter procedure, using the integrability of the graph building operator, as described in this Section.}
\label{cylinderFishnet}}
\end{minipage}
\end{figure}

We start from the case of general $D,w$. Suppose we want to compute the correlator of local operators ${\cal O}_L(z)=\tr[ Z(z)]^L$ in FCFT$_D$ 
\beq\label{eq:globe}
   K(z,y)=\langle {\cal O}_L(z)\,\,\bar {\cal O}_L(y) \rangle\;.
\eeq
Due to the chiral property of the Lagrangian \eqref{bi-scalarL-D} the only  planar  Feynman graphs for this correlator have a ``globe'' configuration Figure \ref{globe_Fishnet}, where ``parallels'' and ``meridians'' form a regular  square-lattice everywhere except for the ``south'' and ``north'' poles. The edges are represented by the propagators in $D$ dimensions given by
\beq
    G_h(x)=(x^{2})^{-\frac{D}{4}+\omega}\;, \quad G_v(x)=(x^{2})^{-\frac{D}{4}-\omega}
\eeq
in horizontal and vertical directions respectively.

According to general properties of CFT, such correlator should have the form 
 \beq\label{eq:corr}
   K(z,y)={\rm const}\,\, |x-y|^{-2\Delta_{{\cal O}_L}}
\eeq
where $\Delta$ is the conformal dimension of this operator.

To compute this correlator, it is worth considering a more general operator,  of the type   ${\cal O}(z_{1}\dots z_{L})=\tr[ Z(z_{1})\dots Z(z_{L})]$ and compute the correlation function 
 \beq\label{eq:KOO}
   K(z_1,\dots,z_L|y_1,\dots,y_L)=\langle {\cal O}(z_{1}\dots z_{L})\,\,\bar {\cal O}(y_{1}\dots y_{L}) \rangle\;.
\eeq
The planar Feynman graphs for such correlator are of cylindrical topology. One should open the ends of meridians converging on the ``south" and ``north" poles on the Figure \ref{globe_Fishnet}). If we then amputate the propagators converging at the south pole we obtain the cylindrical  configuration presented on the Figure \ref{cylinderFishnet}. Such a graph can be represented at each order of perturbation theory as the corresponding power of the so-called graph-building operator, acting on the space $(\mathbb{R}^D)^{\bigotimes L}$,
 \beqa\label{eq:GBO}
H^{(w)}_{L}(z_1,z_2,\dots,z_L|z_1',z_2',\dots,z_L')&=&G_v(x_{11'})G_v(x_{22'})\dots G_v(x_{22'}) \times \notag \\
&\times& G_h(x_{1'2'})G_h(x_{2'3'})\dots G_h(x_{L'1'})\;,
\eeqa
where we use the notations  $x_{ab}=x_{a}-x_{b}$.
Then the above correlator \eqref{eq:KOO} can  be represented as~\cite{Gurdogan:2015csr,Gromov:2017blm} 
\beq
   K(z_1,\dots,z_L|y_1,\dots,y_L)=\langle z_{1}\dots z_{L}|\,\frac{1}{1-\xi^{2L}H^{(s)}_{L}}\,|y_{1}\dots y_{L} \rangle\;,
\eeq
where the states $|y_{1}\dots y_{L} \rangle$ are taken in the usual coordinate representation.
It is easy to see, already by power counting, or applying the inversion transformation, that each cylindrical graph entering the perturbative expansion in $\xi^2$ is conformal. Thus the whole correlator will have the coordinate dependence according to the Euclidean $D$-dimensional conformal symmetry corresponding to the $\mathfrak{so}(1,D+1)$ algebra.

Introducing the 2D momenta conjugated to the coordinates $[p_k,x_k]=i$, we can represent the operator \eqref{eq:GBO} in a more compact form 
\beqa
   H^{(s)}_{L}&=&(p^2_{1})^{-\frac{D}{4}+\omega}(x_{12})^{-\frac{D}{4}-\omega}(p^2_{2})^{-\frac{D}{4}+\omega} \times \\ 
   &\times& (x_{23})^{-\frac{D}{4}-\omega}(p^2_{3})^{-\frac{D}{4}+\omega}\dots (p^2_{L})^{-\frac{D}{4}+\omega}(x_{L1})^{-\frac{D}{4}-\omega}\;. \notag
\eeqa

The operator $H^{(w)}_{L}$ is known to be a conserved charge from the hierarchy of charges of the $so(1,D+1)$ spin chain encoded into the T-operator~\cite{Zamolodchikov:1980mb,Chicherin:2012yn,Kazakov:2018hrh}
\beq\label{eq:Top}
   T_{L}(u)= \Tr_{\textrm{aux}}(\hat R^{1'}_{1}(u)\hat R^{2'}_{2}(u)\dots \hat R^{L'}_{L}(u))
\eeq
built from $L$ $R$-matrices acting in the product of principal series representations of $\mathfrak{so}(1,D+~1)$ with the weights $(\Delta_1,0,0,\dots)\times(\Delta_2,0,0,\dots)$
\beqa\label{eq:R}
\hat R^{2'}_{2}(u) &=&\langle x_1|R^{2'}_{2}(u)|x_1'\rangle = \\
&=& \frac{1}{ (x_{12}^2)^{-u-\frac{D}{2}+\Delta_+}
(x_{21'}^2)^{u+\frac{D}{2}+\Delta_-}
(x_{1'2'}^2)^{-u+\frac{D}{2}-\Delta_+ }
(x_{12'}^2)^{u+\frac{D}{2}-\Delta_- }}\;, \notag
\eeqa
where $\Delta_\pm=\frac{\Delta_1\pm \Delta_2}{2}$. The matrix multiplication in auxiliary space in \eqref{eq:Top} is understood as the integration $\int d^2z\langle x|\dots|z \rangle \langle z|\dots|y \rangle$  and the trace $\Tr_{\textrm{aux}}$  is understood as   $\Tr_{\textrm{aux}}(\dots)=\int d^2z\langle z|\dots|z \rangle$.

We can also use the momenta $p_j$ to represent the $R$-matrix in the form
\beq
  R_{ab}(u,\xi)= (x^2_{ab})^{u+\frac{D}{2}-\Delta_+}(p^2_{a})^{u+\Delta_-}(p^2_{b})^{u-\Delta_-}(x_{ab}^2)^{u-\frac{D}{2}+\Delta_+}\;.
\eeq

Indeed, if we  take $u= - \frac{D}{2}+\Delta_++\epsilon$ and $\Delta_+=\frac{D}{4}$, then the first factor in \eqref{eq:R} in the limit $\epsilon\to 0$ disappears and the last one effectively becomes $\frac{1}{(x_{1'2'})^{\frac{D}{2}}}\sim \epsilon^{-1}\delta^{(D)}(x_{1'2'})$, and it is easy to see  that 
\beq\label{eq:Tlim}
   T_{L}(u)\sim \frac{1}{\epsilon^L} H_L^{(w)}, \quad \text{where} \quad \Delta_1=\frac{D}{4}-w, \quad \Delta_2=\frac{D}{4}+w\;.
\eeq

An interesting particular case of \eqref{eq:Top}, mentioned in \cite{Chicherin:2012yn} is the limit $u\to 0$, when $\Delta_1=\Delta_2\equiv\Delta$, since it is the only way to generate from it the local conserved charge -- the spin chain hamiltonian with nearest neighbours interaction. Indeed, the $R$-matrix becomes in this limit
\beq
  R_{ab}(u,\xi)=1+u h_{ab}+{\cal O}(u^2)\;,
\eeq 
where 
\beqa
  h_{ab}(\Delta)&=& 2\log(x^2_{ab})+(x_{ab}^2)^{\frac{D}{2}-\Delta}
  \log(p^2_{a}p^2_{b})(x_{ab}^2)^{-\frac{D}{2}+\Delta}= \\
   &=& (p_{a}^2)^{-\frac{D}{2}+\Delta}\log(x^2_{ab})(p_{a}^2)^{\frac{D}{2}-\Delta}+(p_{b}^2)^{-\frac{D}{2}+\Delta}\log(x^2_{ab})(p_{b}^2)^{\frac{D}{2}-\Delta}+ \log(p^2_{a}p^2_{b})\;, \notag
\eeqa
which gives the Heisenberg hamiltonian for non-compact $SO(1,D+1)$ spin chain with nearest neighbors interaction of spins (with conformal spin $\Delta$):
\beq
    {\cal H}_\Delta=\sum_{a=1}^L\,  h_{ab}(\Delta)\;,
\eeq
For the particular dimension $D=2$ and conformal spin $\Delta=0$, we obtain from here the famous Lipatov's $SL(2,\mathbb{C})$ spin chain describing the BFKL physics of interacting reggeized gluons in LO approximation~\cite{Lipatov:1993yb}
\beq
  h_{ab}^{BFKL}=2\log(x^2_{ab})+(x_{ab}^2)\log(p^2_{a}p^2_{b})(x_{ab}^2)^{-1}\;.
\eeq
In the particular case of two interacting reggeized gluons (Pomeron state) the spin chain reduces to two-spin hamiltonian
\beq
    {\cal H}_\Delta=2h_{12}^{BFKL}
\eeq
with the spectrum given by the RHS of \eqref{pomeron}.

In this way, we see that the  BFKL physics in LO approximation emerges as a particular case of  $D$-dimensional Fishnet CFT, as it was first noticed in~\cite{Kazakov:2018qez}.

\section{Conclusions}

In this short review we made an attempt to briefly explain the studies of the BFKL spectrum in \sym, which were inspired by the works of Lev Lipatov  on the BFKL integrability in the 4D gauge theories without and with supersymmetry. His ideas have led to a very significant progress in understanding the BFKL spectrum of \sym, where it was possible to build the bridge between the integrability of \sym\ and integrability of the BFKL limit in the gauge theory. The usage of the Quantum Spectral Curve method in this limit has led to a series of new analytic and numerical results mentioned in this review.

The key feature utilized to link the results (including BFKL spectrum) in \sym\ and QCD is the Kotikov-Lipatov's principle of maximal transcendentality \cite{Kotikov:2000pm,Kotikov:2001sc}. In the Section \ref{BFKL_max_transcend} we illustrated this principle with an example of the NLO BFKL kernel eigenvalue in \sym\ and QCD. In general, this principle goes much beyond the BFKL spectrum. Many more observables in \sym\ admit this principle and reproduce  the most transcendental part of the corresponding results in QCD. 
Another key observation by Lipatov~\cite{Kotikov:2000pm}, which allowed to connect the integrability of \sym\ with the BFKL regime, is 
the relation between the BFKL regime and the  dimensions of the operators in \sym\ explained in the Section \ref{dim_analit_cont}.
This enabled the QSC method to be applied for the BFKL spectrum.


We also described here another special limit of \sym\ similar in spirit to the BFKL limit, leading to the Fishnet CFT, currently actively studied in the literature. We explained here the generalization of the Fishnet CFT from four to any number $D$ of dimensions. It appears that, at $D=2$ and a special value of spin of the involved fields, the Fishnet CFT reduces to Lipatov's integrable spin chain describing  the interacting reggeized gluons in LO BFKL approximation of QCD. This remarkable correspondence can open new ways for the study of BFKL physics.

A great deal of the work presented here has been inspired by the ideas of Lev Lipatov. These ideas continue to influence a large community of theoretical physicists working on various non-perturbative aspects of quantum gauge theories. They find applications in a variety of domains of theoretical physics,  including the integrable field theories. The BFKL spectrum of \sym\ is only one of many such applications, where Lev Lipatov's ideas gave rise to a great progress and revealed new intriguing problems, which still wait for their solution.

All three of us had the privilege to know Lev Lipatov personally and had the chance to appreciate, in numerous conversations, his outstanding personality and enormous talent. For one of us (V.K.), Lev Lipatov has been a dear friend over many years.
For another (N.G.), Lev Nikolaevich was a dedicated and caring teacher who, over many years, greatly shaped his taste in research and was always available with deep insight and advice.
And for the latter of us (M.A.), Lev Lipatov, despite being personally acquainted with him for a short time, left a great heritage of inspiring ideas and will always be an outstanding researcher to look up to.

For us, he will always be a great example of selfless devotion to science.

\subsection*{\it Acknowledgements}

M.A. is grateful to Masha N. for her kind support during the work on this text.
The work of N.G. was supported by the ERC grant 865075 EXACTC.

\bibliography{bibliography_Lipatov_mem}
\bibliographystyle{JHEP}


\end{document}